# Posterior inference of attitude-behaviour relationships using latent class choice models

31 March 2026

Akshay Vij
School of Management
Adelaide University
akshay.vij@adelaide.edu.au

Stephane Hess
Choice Modelling Centre
Institute for Transport Studies
University of Leeds
S.Hess@leeds.ac.uk

**Abstract**

The link between attitudes and behaviour has been a key topic in choice modelling for two decades, with the widespread application of ever more complex hybrid choice models. This paper proposes a pragmatic and computationally tractable alternative framework for empirically examining the relationship between attitudes and behaviours using latent class choice models (LCCMs). Rather than embedding attitudinal constructs within the structural model, as in hybrid choice frameworks, we recover class-specific attitudinal profiles through posterior inference. This approach enables analysts to explore attitude-behaviour associations without the complexity and convergence issues often associated with integrated estimation. Two case studies are used to demonstrate the framework: one on employee preferences for working from home, and another on public acceptance of COVID-19 vaccines. Across both studies, we compare posterior profiling of indicator means, fractional multinomial logit (FMNL) models, factor-based representations, and hybrid specifications. We find that posterior inference methods provide behaviourally rich insights with minimal additional complexity, while factor-based models risk discarding key attitudinal information, and full-information hybrid models offer little gain in explanatory power and incur substantially greater estimation burden. Our findings suggest that when the goal is to explain preference heterogeneity, posterior inference offers a practical alternative to hybrid models, one that retains interpretability and robustness without sacrificing behavioural depth.

## 1. Introduction

Understanding the relationship between attitudes and behaviour is critical in fields such as transportation, health, and labour economics, where there is clear scope for decisions to be influenced by complex psychological and perceptual factors. Surveys on behaviour often collect information relating to attitudes, typically in the form of rating scale answers to attitudinal questions. It has long been recognised that the answers to such questions are potentially affected by measurement error and correlated with other unobserved effects, and that their use as covariates in a model puts analysis at risk of endogeneity bias. With a view to avoiding such issues, integrated choice and latent variable (ICLV) models, also known as hybrid choice models (HCMs), have become the gold standard for investigating the relationships between attitudes and behaviour in a more robust manner (Abou-Zeid and Ben-Akiva, 2024; Ben-Akiva et al., 2002a; Ben-Akiva et al., 2002b; Walker, 2001; McFadden, 1986). These models allow for the inclusion of latent constructs, such as attitudes and perceptions, as explanatory variables in discrete choice models by integrating structural equation modelling with choice modelling. This enables researchers to uncover how latent psychological factors are formed and how they shape observed decisions, providing a theoretically robust and behaviourally realistic framework.

However, despite their theoretical appeal, ICLV models come with significant limitations that hinder their practical application (Vij and Walker, 2016; Chorus and Kroesen, 2014). A major challenge lies in their structural complexity, which requires simultaneous estimation of latent variable and discrete choice sub-models (Bahamonde-Birke & de Dios Ortúzar, 2014; Raveau et al., 2010; Walker 2001). This complexity leads to high computational cost, making ICLV models particularly resource-intensive for large-scale or high-dimensional datasets. Additionally, ICLV models often face identification issues, where model parameters cannot be uniquely estimated due to overlapping influences of observed and latent variables (Vij and Walker, 2014). These challenges are compounded by difficulties in interpreting the latent constructs and their estimated relationships, further limiting the accessibility and utility of ICLV models for practitioners and policymakers (Vij and Walker, 2016; Chorus and Kroesen, 2014). Finally, the actual benefits in terms of behavioural insights or prediction performance are often more limited than analysts might expect. Despite these limitations, ICLV models remain the dominant approach for incorporating attitudinal information in applied choice modelling, leaving practitioners with a limited set of practical alternatives when computational constraints, identification challenges, or interpretability concerns arise.

As an alternative to ICLV models, we propose a pragmatic framework based on latent class choice models (LCCMs) (Hess, 2024; Kamakura and Russell, 1989), mitigating the high computational cost and identification issues associated with ICLV models, while enabling transparent analysis of observed attitude–behaviour relationships.. Our approach specifically leverages the posterior probabilities of class membership to profile class-specific mean responses to Likert-scale indicators, which measure attitudes or other latent constructs. We also implement a fractional multinomial logit (FMNL) model that regresses posterior class membership probabilities on attitudinal indicators, allowing for a multivariate analysis of how individual attitudes influence class assignment. Both approaches avoid the need for simultaneous estimation of structural equation and choice sub-models, thereby eliminating model complexity and sidestepping identification issues, without imposing any additional computational costs beyond a standard LCCM. They also offer greater transparency and flexibility, providing interpretable insight into the relationship between observed behaviours and attitudinal heterogeneity. Finally, they avoid the need for analysts to make potentially arbitrary decisions about how attitudinal indicators are grouped into latent constructs, reducing the risk of imposing questionable structure on the data. While we develop the approach with a focus on LCCMs, the same principle can also be used with continuous mixture models, i.e. mixed Logit.

This paper applies the proposed framework to two distinct empirical case studies, demonstrating its versatility and practical value. The first case study examines how worker preferences for working from home (WfH) vary as a function of their perceptions of WfH impacts on productivity, health and wellbeing, and human relations. The second case study examines how individual preferences for COVID-19 vaccines vary as a function of their attitudes such as concern about the pandemic and beliefs about vaccine risks. By profiling class-specific attitudinal responses, we demonstrate the utility of posterior inference in uncovering nuanced attitude-behaviour relationships across diverse contexts.

In summary, this paper makes three key contributions. First, it provides a structured application of posterior inference with LCCMs to investigate attitude–behaviour relationships, addressing practical limitations associated with the estimation and interpretation of ICLV models. Second, it applies this framework to two empirical case studies, illustrating its practical value and versatility across distinct decision contexts. Third, it offers insights into worker preferences for WfH and individual preferences for COVID-19 vaccines, decision contexts with strong implications for transport behaviours, providing actionable evidence for policy analysis and applied decision-making. While each component of the proposed framework draws on established mixture-modelling practices, the contribution of this paper lies in their systematic integration into a unified diagnostic workflow, and in the explicit benchmarking of this workflow against factor-based and hybrid choice specifications across multiple empirical settings, rather than in the development of a new estimator or identification strategy. The approach is particularly suited to settings where attitudinal indicators are closely aligned with the behaviour of interest, and is intended as a transparent diagnostic complement to, rather than a replacement for, more structurally complex hybrid choice models.

The remainder of this paper is structured as follows. Section 2 introduces our proposed framework for inferring attitude-behaviour relationships using posterior inference applied to latent class choice models (LCCMs). It also outlines several benchmark approaches from the existing literature, such as factor-based and hybrid choice models, against which our framework is compared. Section 3 presents the first case study, examining employee preferences for working from home, and demonstrates the proposed posterior inference framework through a series of progressively complex models. Section 4 applies the same modelling sequence to a second case study on public preferences for COVID-19 vaccination, offering a comparative perspective on model performance across different attitudinal structures. Section 5 concludes by summarising key findings, discussing methodological implications, and outlining directions for future research.

## 2. Methodology

This section outlines the methodological framework for posterior inference using latent class choice models (LCCMs) to investigate attitude-behaviour relationships. We describe the key components of the approach, including the estimation of LCCMs, posterior inference of class membership, profiling of class-specific attitudinal responses, and a fractional multinomial logit (FMNL) model to assess the marginal effects of attitudinal indicators on posterior class membership probabilities. We conclude by introducing a set of alternative model specifications, including factor score-based and fully specified ICLV frameworks, which are used to benchmark and contextualise the proposed approach.

### 2.1 Latent Class Choice Model (LCCM) Framework

The LCCM extends traditional discrete choice models by assuming that the population comprises a finite number of latent classes, each characterized by distinct preference structures. The probability of individual $n$ choosing alternative $j$ in choice situation $t$, conditional on latent class $c$, is typically modelled as the following multinomial logit specification:

$$P(y_{nt} = j|c) = \frac{\exp\left(v_{ntj}^{c}\right)}{\sum_{j' \in J} \exp\left(v_{ntj'}^{c}\right)} = \frac{\exp(\mathbf{x}_{\mathbf{nti}}' \boldsymbol{\beta}_{\mathbf{c}})}{\sum_{j' \in J} \exp\left(\mathbf{x}_{\mathbf{ntj'}}' \boldsymbol{\beta}_{\mathbf{c}}\right)} \tag{1}$$

where $v_{ntj}^{c}$ is the utility of alternative $j$ for individual $n$ in class $c$, $J$ is the set of available alternatives, $\mathbf{x_{ntj}}$ is a vector of covariates describing alternative $j$, and $\boldsymbol{\beta}_{\mathbf{c}}$ is a vector of class-specific parameters denoting sensitivities to the same.

The probability of individual $n$ belonging to latent class $c$ is modelled using a class membership model, typically specified also as a multinomial logit model:

$$P(c) = \frac{\exp(\mathbf{z}_{\mathbf{n}}' \boldsymbol{\alpha}_{\mathbf{c}})}{\sum_{\mathbf{c}'} \exp(\mathbf{z}_{\mathbf{n}}' \boldsymbol{\alpha}_{\mathbf{c}'})} \tag{2}$$

where $\mathbf{z_n}$ is a vector of individual-specific covariates, and $\boldsymbol{\alpha}_{\mathbf{c}}$ represents the class-specific coefficients.

The typical assumption is that tastes vary across individuals, but that they are constant for a given individual. The marginal probability of the observed sequence of choices $\mathbf{y_n}$ for person $n$ is then:

$$P(\mathbf{y_n}) = \sum_{c} P(c) P(\mathbf{y_n}|c) = \sum_{c} P(c) \left( \prod_{t} P(y_{nt}|c) \right) \tag{3}$$

Equation (3) can be combined iteratively across individuals in the sample population to derive the following likelihood function:

$$L(\boldsymbol{\alpha}, \boldsymbol{\beta}) = \prod_{n} P(\mathbf{y_n}) \tag{4}$$

The unknown model parameters $\boldsymbol{\alpha}$ and $\boldsymbol{\beta}$ are estimated by maximizing the likelihood function.

### 2.2 Posterior Inference of Class Membership

Once the model has been estimated, we can derive posterior class membership probabilities for each individual $\mathrm{n}$ in the sample population, given their observed choices $\mathbf{y_n}$. Let $\hat{\boldsymbol{\alpha}}$ and $\hat{\boldsymbol{\beta}}$ denote the estimated parameters of the LCCM. The posterior probability that individual $\mathrm{n}$ belongs to class $\mathrm{c}$ can be estimated using Bayes' theorem, as follows:

$$\hat{\mathrm{p}}_{\mathrm{nc}} = \mathrm{P}\left(\mathrm{c}|\mathbf{y_n}; \hat{\boldsymbol{\alpha}}, \hat{\boldsymbol{\beta}}\right) = \frac{\mathrm{P}\left(\mathrm{c}, \mathbf{y_n}|\hat{\boldsymbol{\alpha}}, \hat{\boldsymbol{\beta}}\right)}{\mathrm{P}\left(\mathbf{y_n}|\hat{\boldsymbol{\alpha}}, \hat{\boldsymbol{\beta}}\right)} = \frac{\mathrm{P}\left(\mathrm{c}|\hat{\boldsymbol{\alpha}}, \hat{\boldsymbol{\beta}}\right)\mathrm{P}\left(\mathbf{y_n}|\mathrm{c}; \hat{\boldsymbol{\alpha}}, \hat{\boldsymbol{\beta}}\right)}{\mathrm{P}\left(\mathbf{y_n}|\hat{\boldsymbol{\alpha}}, \hat{\boldsymbol{\beta}}\right)} \tag{5}$$

where $\mathrm{P}\left(\mathbf{y_n}|\mathrm{c}; \hat{\boldsymbol{\alpha}}, \hat{\boldsymbol{\beta}}\right)$ is the likelihood of the observed choices for individual $\mathrm{n}$ conditional on class $\mathrm{c}$, and $\mathrm{P}\left(\mathrm{c}|\hat{\boldsymbol{\alpha}}, \hat{\boldsymbol{\beta}}\right)$ is the prior probability of class membership as given by the class membership model. These posterior probabilities provide a probabilistic assignment of individuals to latent classes, reflecting the degree of belief that an individual belongs to each class given their observed behaviour.

In the subsequent analysis, we treat the posterior class membership probabilities as fixed inputs and condition on the estimated LCCM. For notational simplicity, we omit the hat notation on $\hat{\mathrm{p}}_{\mathrm{nc}}$ in what follows and denote these posterior probabilities as $\mathrm{p}_{\mathrm{nc}}$.

This conditional treatment implies that uncertainty from the estimation of the LCCM is not propagated into the posterior analysis. To assess the robustness of our results to this approximation, we implement a simulation-based procedure in the spirit of Krinsky and Robb (1986) to propagate parameter uncertainty into the posterior probabilities and downstream statistics. Specifically, exploiting the asymptotic normality of the maximum likelihood estimator, we draw repeatedly from the estimated multivariate normal sampling distribution of the LCCM parameters using the estimated covariance matrix, and recompute posterior class membership probabilities and corresponding statistics for each draw. The results are qualitatively unchanged and are reported in Appendix A. These findings indicate that, in this context, treating posterior probabilities as fixed inputs does not materially affect inference, and that the use of point estimates in the posterior analysis provides a reliable and computationally efficient approximation.

### 2.3 Profiling Class-Specific Responses to Attitudinal Indicators

Once posterior class membership probabilities are computed, we profile class-specific mean responses to Likert-scale indicators measuring attitudes or perceptions. Let $\mathrm{i}_{\mathrm{nk}}$ denote individual $\mathrm{n}$'s response to indicator $\mathrm{k}$, and let $\mathrm{p}_{\mathrm{nc}}$ denote the posterior probability that individual $\mathrm{n}$ belongs to class $\mathrm{c}$. The posterior-weighted mean response for indicator $\mathrm{k}$ in class $\mathrm{c}$ is given by:

$$\hat{\mu}_{\mathrm{kc}} = \mathrm{E}[\mathrm{i}_{\mathrm{k}}|\mathrm{c}] = \frac{\sum_{\mathrm{n}} \mathrm{p}_{\mathrm{nc}} \cdot \mathrm{i}_{\mathrm{nk}}}{\sum_{\mathrm{n}} \mathrm{p}_{\mathrm{nc}}} \tag{6}$$

To assess whether differences in mean responses across classes are statistically significant, we first conduct a one-way ANOVA test for each indicator. The ANOVA tests the null hypothesis that the class-specific mean responses $\hat{\mu}_{\mathrm{kc}}$ are equal across all classes. The ANOVA test statistic is given by the following weighted F-statistic:

$$\mathrm{F} = \frac{\sum_{\mathrm{c}}\left(\sum_{\mathrm{n}} \mathrm{p}_{\mathrm{nc}}\right)\left(\hat{\mu}_{\mathrm{kc}} - \mathrm{E}[\mathrm{i}_{\mathrm{k}}]\right)^2}{\sum_{\mathrm{c}} \sum_{\mathrm{n}} \mathrm{p}_{\mathrm{nc}} \left(\mathrm{i}_{\mathrm{nk}} - \hat{\mu}_{\mathrm{kc}}\right)^2} \cdot \frac{\mathrm{N} - \mathrm{C}}{\mathrm{C} - 1} \tag{7}$$

where $E[i_k]$ is the overall sample mean for indicator $k$; $N$ is the sample size, $C$ denotes the number of classes, and $F$ is the ANOVA test statistic which follows an F-distribution with $C-1$ and $N-C$ degrees of freedom under the null hypothesis that all class means are equal.

If the ANOVA test indicates statistically significant within-class differences for one or more indicator responses, we can conduct follow-up t-tests to identify statistically significant differences between particular pairs of classes. Because both class-specific means are computed from the same sample of individuals using posterior weights, they are not independent. We therefore compute the corresponding t-statistic using a covariance-adjusted variance estimator. For any pair of classes $c$ and $c'$, the t-test statistic for indicator $k$ is given by:

$$t = \frac{\hat{\mu}_{kc} - \hat{\mu}_{kc'}}{\sqrt{\widehat{Var}(\hat{\mu}_{kc}) + \widehat{Var}(\hat{\mu}_{kc'}) - 2\widehat{Cov}(\hat{\mu}_{kc}, \hat{\mu}_{kc'})}} \quad (8)$$

, where the posterior-weighted variance and covariance estimates are given by:

$$\widehat{Var}(\hat{\mu}_{kc}) = \frac{\sum_n p_{nc}^2 \cdot (i_{nk} - \hat{\mu}_{kc})^2}{(\sum_n p_{nc})^2} \quad (9)$$

$$\widehat{Cov}(\hat{\mu}_{kc}, \hat{\mu}_{kc'}) = \frac{\sum_n p_{nc} p_{nc'} (i_{nk} - \hat{\mu}_{kc})(i_{nk} - \hat{\mu}_{kc'})}{(\sum_n p_{nc})(\sum_n p_{nc'})} \quad (10)$$

The ANOVA provides a global test for each indicator, identifying whether class-specific means differ significantly across all classes. In contrast, the pairwise t-tests are used to localise these differences, identifying which specific class pairs exhibit statistically significant differences. While the large number of pairwise comparisons introduces a risk of inflated Type I error, we do not apply formal corrections such as Bonferroni or Holm adjustments. This is because the ANOVA and t-tests serve distinct, complementary purposes in our analysis: the former provides a global test of heterogeneity, while the latter supports interpretive clarity by illustrating where behavioural differences lie. As our goal is exploratory rather than confirmatory hypothesis testing, we report results transparently and encourage contextual interpretation of statistical significance. The pairwise comparisons are therefore interpreted as post-hoc diagnostics to localise differences identified by the global ANOVA test, rather than as a set of independent hypothesis tests.

The statistical testing framework allows us to identify which attitudinal differences between classes are statistically significant, providing deeper insights into the heterogeneity of attitudes and their relationship with observed behaviours. The same approach can also be applied to factor scores, enabling the analyst to profile classes in terms of differences in underlying latent constructs, when indicators have been grouped through factor analysis, thus applying Equations (6)-(10) to factor scores rather than to individual attitudinal statements. An analyst can also still make links with observed decision maker characteristics by applying Equation (6) to such variables and thus creating a profile for the socio-demographic composition of a class, and studying the correlation between the socio-demographic and attitudinal profile of each class.

### 2.4 Fractional Multinomial Logit Model

The approach in Section 2.3 focusses on one attitudinal question at a time. The analyst can further treat the posterior class membership probabilities derived from a baseline latent class choice model (LCCM) as the dependent variable and use the attitudinal indicators as covariates in a fractional multinomial logit (FMNL) model. The FMNL model assumes the following structure for each individual $\mathrm{n}$ and class $\mathrm{c} \in \{2, \ldots, \mathrm{C}\}$, where $\mathrm{C}$ denotes the total number of classes:

$$\mathrm{P}(\mathrm{c}|\mathbf{y_n}) = \frac{\exp(\mathbf{i}'_\mathbf{n}\boldsymbol{\gamma}_\mathbf{c})}{\sum_{\mathrm{c}'} \exp(\mathbf{i}'_\mathbf{n}\boldsymbol{\gamma}_{\mathbf{c}'})} \tag{11}$$

, where $\boldsymbol{\gamma}_\mathbf{c}$ is the vector of parameters capturing the association between the attitudinal indicators and posterior class membership probabilities associated with class $\mathrm{c}$, relative to the probability that the individual belongs to the reference class. The FMNL model is estimated by maximum likelihood using the multinomial logit functional form applied to fractional outcomes, as given by (11), ensuring that predicted class membership probabilities lie in the unit interval and sum to one across classes.

The FMNL approach provides an alternative way to examine the relationship between attitudes and latent class membership. Whereas the posterior profiling method described in Section 2.3 relies on univariate comparisons of class-specific means and variances for each indicator, the FMNL approach adopts a multivariate perspective. It allows us to quantify the marginal association between each indicator and class membership while accounting for the potential confounding influence of other indicators. As with the posterior profiling approach, the FMNL approach can also be applied using factor scores as explanatory variables, offering a way to explore how variation in underlying latent constructs is associated with class membership while accounting for correlations between related indicators.

The FMNL model thus enables a more formal diagnostic test of the associations observed in the posterior summaries, offering a valuable middle ground between informal posterior inference and more complex structural modelling. This method maintains the class definitions and choice model parameters fixed, avoiding the circularity, identification and endogeneity concerns associated with other analogous approaches, such as the direct inclusion of indicators within the choice model[1]. It also retains many of the advantages of the posterior inference framework, such as computational simplicity, transparency, and behavioural interpretability, while enabling more rigorous statistical testing of the underlying attitudinal relationships. Importantly, since posterior class membership probabilities are themselves derived from estimated choice-model parameters, the FMNL coefficients should be interpreted as descriptive diagnostics of association rather than structural behavioural parameters.

### 2.5 Comparisons with Alternative Frameworks

Over the following sections, we use two case studies to demonstrate how the proposed inference framework can be applied in practice to uncover meaningful relationships between attitudes and behaviours. For each case study, we estimate a baseline LCCM, compute posterior class membership probabilities, profile class-specific attitudinal responses using the posterior profiling framework outlined in Section 2.3, and estimate FMNLs using the approach described in Section 2.4. To benchmark the proposed frameworks, we compare them against a range of alternative modelling strategies commonly used to examine the relationship between attitudes and behaviour.

---

[1] While posterior inference avoids simultaneity and feedback between attitudinal constructs and choice behaviour that arise under joint estimation, it does not address other potential sources of endogeneity due to unobserved traits or omitted variables that may jointly influence attitudes and choices, nor does it eliminate concerns related to measurement error in attitudinal indicators.

The first of these is the widely used integrated choice and latent variable (ICLV) model, where in our case, the latent variables are used to explain class membership. The ICLV model jointly estimates latent attitudinal constructs and behavioural choices through a fully integrated structural framework. It models responses to attitudinal indicators via a measurement model, links these constructs to class membership through a structural equation model, and estimates all components simultaneously. This approach is the gold standard in the field, due to its theoretical robustness in addressing measurement error, endogeneity, and latent structure. However, as discussed later, this rigour often comes with considerable practical costs.

We also include two pragmatic approaches that incorporate attitudinal data directly into the estimation of class membership, bypassing the measurement model altogether. The first of these includes attitudinal indicators directly in the class membership model (Model 2), while the second first applies factor analysis to collapse the indicators into a smaller number of latent scores, which are then used as explanatory variables (Model 3). These models are easier to estimate and interpret, and avoid the convergence and identification issues that often afflict ICLV models. However, Model 2 is susceptible to endogeneity bias, since the attitudinal indicators may be jointly determined with the choice outcomes through unobserved common causes. Model 3, by using factor scores that represent the underlying latent constructs, is designed to address this source of bias. However, Model 3 introduces a different issue: the factor scores are treated as if they were directly observed without error, thereby ignoring the estimation variance from the measurement model. This "errors-in-variables" problem can lead to attenuation of estimated effects. In practice, the two models thus address different concerns, but neither is free of potential bias.

To further test the robustness of our findings and evaluate the explanatory power of attitudinal indicators without modifying the underlying behavioural segmentation, we estimate two additional benchmark models. These are *sequential* LCCMs in which the class-specific choice model parameters are fixed at their baseline estimates from the initial LCCM, and only the class membership model is re-estimated using either the attitudinal indicators or the factor scores as explanatory variables. While these specifications mirror Model 2 and Model 3 in their use of indicator-based and factor-based inputs, respectively, they differ in a crucial way: by preserving the original segmentation structure and eliminating feedback between the measurement and choice components, they serve as cleaner diagnostic tools. In this respect, they are closer in spirit to the FMNL approach introduced in Section 2.4, and help isolate the explanatory contribution of attitudinal information.

To situate the variety of model structures described above, **Figure 1** illustrates the relationship between attitudes and behaviours as modelled by various indicator-based approaches, while **Figure 2** illustrates the same for various factor-based approaches. Taken together, these schematics underscore how the alternative strategies we review provide a robust and comprehensive basis for evaluating the relative strengths and trade-offs of different approaches to modelling attitude–behaviour relationships.

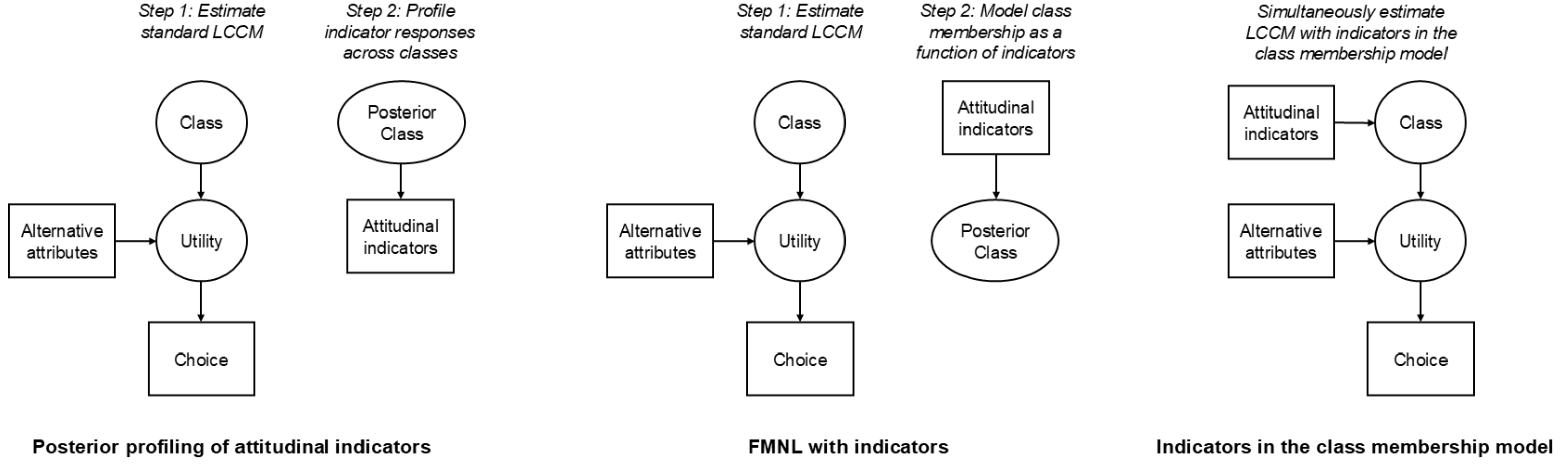

**Figure 1:** Schematic showing how different indicator-based model structures and specifications capture the relationship between attitudes and behaviours

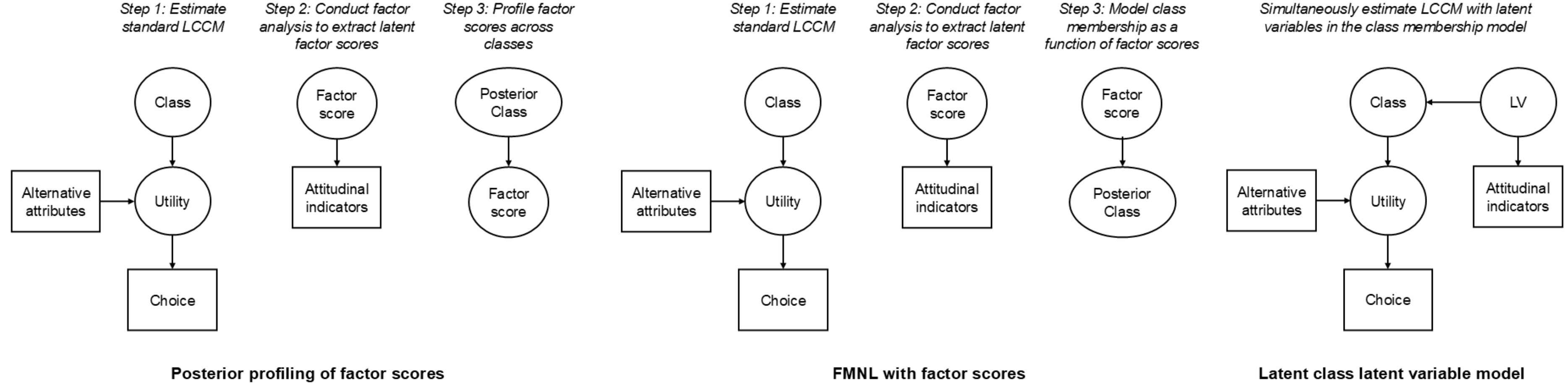

**Figure 2:** Schematic showing how different factor-based model structures and specifications capture the relationship between attitudes and behaviours

## 3. Case Study 1: Employee Preferences for Working from Home (WfH)

The first case study uses the proposed framework to examine how employee preferences for WfH might vary as a function of perceived impacts of WfH on productivity, health and wellbeing, and human relations.

Data for our analysis comes from Vij et al. (2023). The dataset comprises responses from 996 employees surveyed in 2020–21, drawn from the 17 largest urban areas in Australia. Each participant had a designated workplace that they worked from or reported to, and indicated that some of their jobs' tasks and activities could be done remotely (if appropriate policies and resources were in place). Survey participants were asked about their current job, their ability to work remotely given the characteristics of their job, and potential uptake of remote working arrangements if they were offered the opportunity to work remotely whenever possible.

The survey included stated preference (SP) experiment scenarios to elicit participants' preferences for different remote working arrangements for themselves, such as the example scenario shown in **Figure 3**. Each respondent was shown 8 scenarios, and the job attributes were varied systematically across scenarios and the values listed in **Table 1**, based on a fractional orthogonal design. To populate the 'Yearly (weekly) take home pay after tax' value, we took a two-tiered sample approach. Firstly, because the individual respondent's current wage was known, five salary ranges were developed as a percentage of current wage rate (these percentage ranges are shown as attribute 3 in **Table 1**). For each scenario that was presented to the respondent, one of these ranges was randomly selected, and from within that range a salary amount was generated. For example, in **Figure 3**, these generated amounts are shown as $103,220 per year and $107,120 per year. For more details about the survey design and data collection, please refer to Vij et al. (2023).

Data from the SP scenarios was used in conjunction with other employment and demographic information collected as part of the survey to estimate LCCMs of employee preferences for remote working. We estimated a number of LCCMs with different model specifications, where we varied the explanatory variables, the functional form of the utilities, and the number of classes. Based both on statistical measures of fit and behavioral interpretation, we select the four-class LCCM as the preferred model specification. For the sake of brevity, we do not include any further details on the model selection process; the interested reader is referred to Vij et al. (2023) for more information. All models for this study were estimated using the software package Apollo (Hess and Palma, 2019).

The final four-class model specification has a McFadden's adjusted R-squared of 0.347, indicating reasonable goodness-of-fit. The model specification, class enumeration, and estimation diagnostics are reported in detail in the original study (Vij et al., 2023). In brief, a range of LCCM specifications were estimated by varying the number of classes, explanatory variables, and functional forms. Model selection was based on a combination of statistical fit (log-likelihood, AIC, BIC) and behavioural interpretability. While a five-class model achieved the lowest AIC, the four-class model was preferred as it provided the lowest BIC and the most behaviourally meaningful segmentation. Estimation was conducted incrementally, with simpler models used to generate starting values for more complex specifications (e.g. the two-class model informing the three-class model, and so on), reflecting the well-known sensitivity of LCCMs to initial conditions. Multiple random starting values were also used to mitigate local maxima concerns. The resulting solutions were found to be stable across runs and consistent across different estimation platforms (Biogeme and Apollo). To reflect the assumption that, all else being equal, workers should prefer jobs that offer greater flexibility and higher wages, the coefficients on the WfH and wage attributes in the class-specific choice models were constrained to be non-negative[2]. For some classes, these parameters reached the zero bound and were consequently not estimated. These constraints were tested as part of the model development process and were found not to materially affect the estimated class structure or posterior class membership probabilities.

For the sake of model parsimony, the class membership model was specified as a constants-only model, and did not include any employment or demographic characteristics as explanatory variables. The class-specific choice models included the three attributes shown in the SP experiments, namely ability to work remotely some days and hours, and wages, as the explanatory variables $\mathbf{x_{ntj}}$. Corresponding estimates for the model parameters $\boldsymbol{\beta}_\mathbf{c}$ are shown in **Table 2**, and a summary of the classes in terms of their shares in the sample population and their compensating wage differentials for the ability to work from home is reported in Table 3.

Note that to ease interpretation and readability, the classes have been ordered in terms of their increasing valuation of remote working arrangements. In summary, roughly half of the sample population (belonging to Classes 1 and 2) does not value the ability to work from home some workdays and/or workhours, while the other half (belonging to Classes 3 and 4) do ascribe a positive and statistically significant value to the same.

---

[2] The non-negativity constraint on the work-from-home attribute reflects an assumption of weak preference for flexibility rather than for remote work per se. In the stated choice scenarios, the attribute indicates whether the worker is permitted to work from home, not whether they are required to do so, nor whether others in the firm do so. A job that allows work from home therefore weakly dominates one that does not, as workers who prefer in-person work may still attend the workplace full-time. Potential disutility arising from others' remote work (e.g., impacts on meetings or workplace interaction) is not explicitly represented in the experiment and is outside the scope of the modelling framework adopted here.

Imagine that the COVID-19 pandemic has ended and the disease has been eradicated.

In this scenario, imagine further that your employer offers you a choice between the following two work arrangements that differ in terms of their flexibility and wages.

Which would you prefer?

**Scenario 1 of 8**

| | **Offer 1** | **Offer 2** |
|---|---|---|
| **Flexibility to work remotely on some days** | **No**, you need to be on-site on all workdays | **Yes**, when possible, you can choose to work some of your workdays remotely |
| **Flexibility to work remotely at some hours** | **Yes**, when possible, you can choose to work some of your work hours remotely | **No**, on the days that you need to be on-site, you need to be on-site at all work hours |
| **Yearly (weekly) take home pay after tax** | $103,220 per year ($1,985 per week) | $107,120 per year ($2,060 per week) |
| **Offer I prefer:** | ○ | ○ |

**Figure 3:** Example screenshot of hypothetical stated preference (SP) scenario to elicit employee preferences for remote and flexible working arrangements

**Table 1:** Range of attribute values used in our stated preference (SP) experiments to describe different working arrangements across different scenarios

| # | Attribute | Range of values |
|---|---|---|
| 1 | Flexibility to work remotely on some days | Yes, when possible, you can choose to work some of your workdays remotely |
| | | No, you need to be on-site on all workdays |
| 2 | Flexibility to work remotely at some hours | Yes, when possible, you can choose to work some of your work hours remotely |
| | | No, on the days that you need to be on-site, you need to be on-site at all workhours |
| 3 | Yearly (weekly) take home pay after tax | Pay between -25% and -15% of current wage rate |
| | | Pay between -15% and -5% of current wage rate |
| | | Pay between -5% and +5% of current wage rate |
| | | Pay between +5% and +15% of current wage rate |
| | | Pay between +15% and +25% of current wage rate |

**Table 2:** Class-specific choice models of employee preferences for remote working

| Variable | Class 1 | | Class 2 | | Class 3 | | Class 4 | |
|---|---|---|---|---|---|---|---|---|
| | **est.** | **p-value** | **est.** | **p-value** | **est.** | **p-value** | **est.** | **p-value** |
| Able to work remotely some days, when possible | 0.317 | 0.19 | 0.000 | - | 2.059 | 0.00 | 3.062 | 0.00 |
| Able to work remotely some hours, when possible | 0.000 | - | 0.028 | 0.83 | 1.171 | 0.00 | 1.475 | 0.00 |
| Wages ($1,000) | 1.142 | 0.00 | 0.006 | 0.34 | 0.493 | 0.00 | 0.119 | 0.00 |

**Table 3:** Summary statistics of 4-class model of employee preferences for remote working

| Attribute | Class 1 | Class 2 | Class 3 | Class 4 | Sample mean | Sample median |
|---|---|---|---|---|---|---|
| Share of the sample population | 29.2% | 24.6% | 26.0% | 20.1% | - | - |
| ***Compensating wage differentials*** | | | | | | |
| Able to work remotely some workdays, when possible | $0* | $0* | $4,174 | $25,731 | $7,526 | $4,078 |
| Able to work remotely some workhours, when possible | $0* | $0* | $2,374 | $12,395 | $3,698 | $2,267 |

** Compensating wage differentials are set to zero in cases where the corresponding taste parameter in the utility function is not statistically significant at the 5 per cent level*

### 3.1 Model 1: Posterior Profiling and the FMNL model

The survey instrument collected responses to a number of Likert-scale statements seeking to measure perceived impacts of working from home on productivity, health and wellbeing, and human relations. We begin by applying the posterior profiling approach to examine whether and how these perceived impacts vary across the four latent classes identified in the baseline LCCM. **Table 4** compares the mean responses to the indicators across different classes.

To begin, we conduct a one-way ANOVA test to assess whether the mean posterior expectations for the responses to each attitudinal indicator vary significantly across the four latent classes identified by our baseline specification. Results reveal that for all indicators, the class-specific means differ significantly at the 0.001 level, suggesting that class membership is strongly associated with systematic variation in attitudes. The relative size of the F-statistic across indicators offers additional insight into which attitudinal constructs contribute most to class differentiation. The lowest F-values are observed for indicators related to perceived impacts on productivity, implying that while statistically significant, differences across classes are less pronounced along this dimension. Higher F-values are observed for indicators measuring perceived impacts on health and wellbeing, and the highest for those relating to human relations, suggesting that these constructs play a particularly important role in distinguishing between the attitudinal profiles of each class.

Next, we examine differences between different subsets of classes, using the pairwise t-test statistics. First, we compare Class 1 to Classes 3 and 4. We find that Class 1 perceives fewer benefits in terms of productivity or health and wellbeing, and the difference is statistically significant across all indicators. This likely explains why they do not value the ability to work from home (c.f. **Table 3**). We also observe some differences in mean responses to indicators measuring perceived impacts on human relations, but these differences are smaller, and statistically insignificant in most cases, indicating that this is likely a less important factor.

Next, we compare Class 2 to Classes 3 and 4. In terms of indicators measuring perceived impacts on productivity and health and wellbeing, we observe small differences between the classes. Mean responses to some measurement indicators are indeed statistically significantly different, but there is no clear consistent trend. However, when we examine mean responses to indicators measuring perceived impacts on human relations, we observe a much clearer, and more statistically significant, difference between the classes. In particular, Class 2 seems to have greater concerns around negative impacts on human relations across all indicators, compared to Classes 3 and 4, explaining why they do not value the ability to work from home (c.f. **Table 3**).

We run an FMNL model in which the posterior class membership probabilities from the baseline four-class LCCM serve as the dependent variable, and the attitudinal indicators are used as explanatory variables (with Class 1 being the reference class). The estimation results are reported in **Table 5**.

**Table 4:** Comparison between mean responses to different attitudinal statements across classes using the baseline LCCM

| Attitudinal construct | **Measure** *(Level of agreement with statements about self: 1 – strongly disagree, 7 – strongly agree)* | **Mean value** | | | | **ANOVA F-stat** [a] | **t-stat** | | | | | |
|---|---|---|---|---|---|---|---|---|---|---|---|---|
| | | | | | | | **Class 1 v.** | | | **Class 2 v.** | | **Class 3 v.** |
| | | **Class 1** | **Class 2** | **Class 3** | **Class 4** | | **Class 2** | **Class 3** | **Class 4** | **Class 3** | **Class 4** | **Class 4** |
| Perceived impacts on productivity | I would be able to focus better on my work | 4.78 | 5.27 | 5.07 | 5.41 | 9.04 *** | 4.22*** | 2.87** | 4.99*** | 2.08* | 1.21 | 3.26** |
| | I would be able to achieve my job objectives and outputs as expected | 4.97 | 5.28 | 5.35 | 5.61 | 8.70 *** | 2.70** | 4.05*** | 5.46*** | 0.68 | 2.93** | 2.64** |
| | I would have an increased sense of self-discipline | 4.68 | 5.16 | 4.91 | 5.19 | 6.91 *** | 4.07*** | 2.42* | 4.24*** | 2.33* | 0.31 | 2.66** |
| | I would be able to multi-task more effectively | 4.74 | 5.10 | 4.90 | 5.37 | 8.60 *** | 3.08** | 1.68 | 5.34*** | 1.94 | 2.43* | 4.75*** |
| | | | | | | | | | | | | |
| Perceived impacts on health and wellbeing | I would have greater life satisfaction | 4.73 | 5.18 | 5.22 | 5.49 | 11.98 *** | 3.97*** | 5.23*** | 6.21*** | 0.39 | 2.61** | 2.44* |
| | I would have higher morale | 4.42 | 5.08 | 4.73 | 5.13 | 12.84 *** | 5.53*** | 3.10** | 5.77*** | 3.27*** | 0.44 | 3.72*** |
| | I would have better work-life balance | 4.91 | 5.22 | 5.37 | 5.76 | 13.20 *** | 2.57** | 4.51*** | 7.12*** | 1.35 | 4.84*** | 3.89*** |
| | I would experience less stress | 4.57 | 5.08 | 4.81 | 5.17 | 7.48 *** | 4.17*** | 2.26* | 4.52*** | 2.43* | 0.67 | 3.01** |
| | | | | | | | | | | | | |
| Perceived impacts on human relations | I would have access to fewer learning opportunities and training sessions | 4.17 | 4.89 | 3.87 | 3.49 | 27.67 *** | 5.65*** | 2.55** | 4.61*** | 8.52*** | 10.17*** | 2.86** |
| | I would be concerned about how my performance would be monitored and observed | 4.25 | 4.93 | 4.08 | 3.82 | 18.57 *** | 5.56*** | 1.45 | 2.92** | 7.21*** | 8.02*** | 1.94 |
| | I would be worried that my colleagues are not doing their fair share of the work | 3.95 | 4.71 | 3.68 | 3.52 | 20.11 *** | 5.64*** | 2.24* | 2.75** | 8.36*** | 8.22*** | 1.14 |
| | The relationship with my supervisor would be adversely affected | 3.85 | 4.69 | 3.57 | 3.32 | 29.19 *** | 6.52*** | 2.46* | 3.68*** | 9.36*** | 10.01*** | 1.91 |
| | My career prospects may suffer due to loss of ad-hoc interactions with colleagues and supervisors | 4.14 | 4.79 | 4.02 | 3.80 | 14.70 *** | 5.15*** | 1.03 | 2.36* | 6.35*** | 7.11*** | 1.69 |

[a] *For an F-distribution with (3, 992) degrees of freedom, if F > 2.60, p-value < 0.05; if F > 3.84, p-value < 0.01; and if F > 6.68, p-value < 0.001*

** p-value < 0.05; ** p-value < 0.01; and *** p-value < 0.001*

**Table 5:** Fractional logit model when the posterior class membership probabilities from the baseline LCCM are the dependent variables

| Variable | Measure<br>*(Level of agreement with statements about self: 1 – strongly disagree, 7 – strongly agree)* | Class 1 (reference) | | Class 2 | | Class 3 | | Class 4 | |
|---|---|---|---|---|---|---|---|---|---|
| | | est. | p-value | est. | p-value | est. | p-value | est. | p-value |
| Class-specific constant | NA | 0.000 | - | -3.052 | 0.00 | -1.003 | 0.00 | -2.248 | 0.00 |
| Perceived impacts on productivity | I would be able to focus better on my work | 0.000 | - | 0.117 | 0.14 | 0.003 | 0.96 | 0.005 | 0.95 |
| | I would be able to achieve my job objectives and outputs as expected | 0.000 | - | 0.022 | 0.77 | 0.127 | 0.03 | 0.081 | 0.33 |
| | I would have an increased sense of self-discipline | 0.000 | - | 0.011 | 0.90 | 0.013 | 0.84 | 0.024 | 0.76 |
| | I would be able to multi-task more effectively | 0.000 | - | -0.076 | 0.35 | -0.114 | 0.09 | 0.084 | 0.34 |
| Perceived impacts on health and wellbeing | I would have greater life satisfaction | 0.000 | - | 0.036 | 0.69 | 0.214 | 0.00 | 0.063 | 0.54 |
| | I would have higher morale | 0.000 | - | 0.220 | 0.01 | -0.016 | 0.83 | 0.096 | 0.27 |
| | I would have better work-life balance | 0.000 | - | -0.088 | 0.27 | 0.067 | 0.30 | 0.207 | 0.02 |
| | I would experience less stress | 0.000 | - | 0.020 | 0.77 | -0.050 | 0.42 | -0.028 | 0.71 |
| Perceived impacts on human relations | I would have access to fewer learning opportunities and training sessions | 0.000 | - | 0.090 | 0.13 | -0.065 | 0.26 | -0.163 | 0.02 |
| | I would be concerned about how my performance would be monitored and observed | 0.000 | - | 0.052 | 0.43 | -0.005 | 0.93 | -0.053 | 0.45 |
| | I would be worried that my colleagues are not doing their fair share of the work | 0.000 | - | 0.052 | 0.38 | -0.047 | 0.36 | -0.032 | 0.62 |
| | The relationship with my supervisor would be adversely affected | 0.000 | - | 0.179 | 0.01 | -0.037 | 0.54 | -0.015 | 0.84 |
| | My career prospects may suffer due to loss of ad-hoc interactions with colleagues and supervisors | 0.000 | - | -0.001 | 0.99 | 0.050 | 0.40 | 0.035 | 0.63 |

As before, we begin by comparing Class 1 with Classes 3 and 4. Consistent with our previous findings, we observe that Class 1 perceives fewer productivity benefits than Class 3 ("I would be able to achieve my job objectives and outputs as expected") and, to a lesser extent, Class 4 as well. Similarly, we observe that Class 1 perceives fewer health and wellbeing benefits than Classes 3 and 4, and the difference is statistically significant across multiple indicators ("I would have greater life satisfaction" and "I would have better work-life balance"). Finally, Class 1 also perceives greater human relations issues than Classes 3 and 4 ("I would have access to fewer learning opportunities and training sessions").

We observe similar trends as before between Classes 2, 3 and 4. Perceived impacts on productivity and health and wellbeing have some impact on class membership, but the trend is not always consistent. For example, compared to Class 2, both Classes 3 and 4 are more likely to agree that they would be able to achieve their job objectives and outputs as expected when working from home. However, the difference is only statistically significant between Classes 2 and 3, and *not* Classes 2 and 4. Similar observations can be made for other indicators. For example, Class 2 is most likely to believe that they "would have higher morale", more so than Classes 3 and 4, and the difference is statistically significant for both. However, Classes 3 and 4 are more likely to believe they "would have better work-life balance", but the difference is statistically significant only between Classes 2 and 4.

In contrast, Class 2 is more likely to agree with all five indicators measuring perceived negative impacts on human relations, compared to Classes 3 and 4, and the impacts are statistically significant for two of these indicators ("I would have access to fewer learning opportunities and training sessions" and "The relationship with my supervisor would be adversely affected"). Once we account for responses to these two indicators, we find that responses to the other indicators do not seem to have a statistically significant impact on class membership (even though our posterior analysis found statistically significant differences in posterior means for all five indicators). The FMNL approach allows us to assess which associations remain robust when multiple correlated indicators are considered jointly.

Conversely, one could argue that the FMNL approach is constrained by the challenge of multicollinearity. Many of the attitudinal indicators used in the FMNL model to explain class membership capture overlapping dimensions of the broader WfH experience, namely perceived impacts on productivity, health and wellbeing, and human relations, and are therefore strongly correlated - multicollinearity is expected. As a result, it becomes difficult to statistically isolate the unique effect of each individual indicator on class membership, which likely contributes to the lack of significance for several parameters. However, this does not invalidate the model. Rather, it reflects the difficulty of isolating marginal effects in the presence of correlated explanatory variables, and should be interpreted as part of the diagnostic value of the approach.

In contrast, the posterior profiling method avoids the need for joint estimation of correlated indicators and thus provides a clearer descriptive account of class-level attitudinal patterns. However, it does not account for the confounding influence of other indicators when interpreting these patterns, and risks overstating the impact of individual indicators. The FMNL and posterior profiling approaches thus offer complementary perspectives - one emphasizing statistical control and marginal effects in a multivariate setting, the other prioritizing transparency and descriptive clarity - allowing analysts to choose the framework best suited to their specific research goals and data characteristics.

### 3.2 Model 2: Indicators in the Class Membership Model

Model 2 involves the simultaneous estimation of a latent class choice model in which attitudinal indicators directly enter the class membership model (right panel of Figure 1). This integrated approach contrasts with the FMNL model discussed in Section 3.1, which holds the underlying class segmentation fixed. The estimation results for the model are reported in Appendix B; please note that the class-specific choice models were found to be nearly identical to the baseline LCCM. To test the robustness of our findings, we also estimate a sequential LCCM in which the class-specific choice model parameters are constrained to the estimates from the baseline LCCM, and only the class membership model is re-estimated using the attitudinal indicators as covariates. Like the FMNL model, this specification preserves the latent class structure while examining how strongly the indicators are associated with class assignment, allowing for a focused investigation of associations without altering the underlying behavioural segmentation. The estimation results are reported in Appendix B. In terms of the magnitude and directionality of effects, both models produce estimation results nearly identical to the FMNL approach, and for the sake of brevity we do not describe them in detail again.

Some critics have argued that the direct inclusion of attitudinal indicators as explanatory variables introduces risks of endogeneity. This concern arises from the possibility that both choice and indicator responses may be jointly determined by latent factors, such as underlying attitudes and perceptions, which are not accounted for explicitly in the model. Alternatively, it may be the case that indicator responses are themselves shaped by prior choices, such that using them to explain current choices risks introducing reverse causality (Chorus and Kroesen, 2014). In either case, the standard exogeneity assumption is violated, and the resulting parameter estimates may be biased or inconsistent. While these are valid concerns in contexts where such feedback loops or confounding influences are likely, we believe that the issue has often been overstated, particularly outside the narrow theoretical contexts in which it was originally raised (see, for example, Ben-Akiva et al., 2002b).

The development of hybrid choice models was partly motivated by a desire to address potential endogeneity bias when attitudinal indicators were used to explain observed choices. However, it is important to recognize that all models, including ICLVs, ultimately estimate statistical associations, not causal effects. Whether a relationship is interpreted as causal depends entirely on the analyst's assumptions and the underlying behavioural theory. If the analyst has a reasoned basis to believe that variation in attitudes (as captured by indicators) explains variation in choices or class membership, then including such indicators directly is a statistically valid and interpretable approach. Conversely, if there is strong reason to believe that indicators are themselves determined by the outcomes of interest, or confounded by omitted variables, then more elaborate structural models like ICLVs may be justified. There is no universal rule that applies across all contexts. In our case, the assumption that individuals' preferences for flexible work arrangements are shaped by their perceptions of WfH impacts on productivity, wellbeing, and human relations is theoretically sound and behaviourally plausible. On that basis, we treat the attitudinal indicators as explanatory variables in the class membership model.

Empirically, our findings reveal that these theoretical concerns have limited practical consequences in this application. The direct inclusion of attitudinal indicators in the class membership model under simultaneous estimation (Model 2) produces estimation results that are nearly identical to those obtained using the FMNL approach and the sequential LCCM, both of which preserve the original class segmentation and are arguably behaviourally more defensible. Despite the differing assumptions about the causal structure between choices and indicators, the patterns of association remain remarkably consistent across all three approaches. This consistency reinforces our broader point that, while concerns about endogeneity are not without merit, their impact may be overstated in much of the literature. In many practical applications, including the one at hand, these modelling choices appear to matter far less than is often assumed, and simpler or more transparent approaches may offer equally valid insights without added complexity.

### 3.3 Model 3: Factor Scores in the Class Membership Model

Next, we conduct a factor analysis to derive factor scores for each of the three latent variables of interest. While the attitudinal indicators were developed with clear domain relationships in mind, broadly aimed at capturing perceived impacts on health and wellbeing, productivity, and human relations, our exploratory factor analysis confirmed this intended structure. The indicators loaded cleanly onto the three expected dimensions, supporting their validity and reliability as measures of the underlying latent constructs. We include these scores as observable variables in the class membership model (Model 3; similar to right panel of Figure 2, but with sequential estimation). The results for the class membership model are reported in **Table 6**. Estimation results for the class-specific choice models are reported in Appendix B; these were found to be nearly identical to the baseline LCCM.

As before, we begin by comparing Class 1 with Classes 3 and 4. Perceived impacts on productivity are not found to have a statistically significant impact on class membership. Classes 3 and 4 both perceive greater health and wellbeing benefits, and fewer negative human relations impacts, than Class 1, and the effect is statistically significant in both cases. Between Classes 2, 3 and 4, perceived negative impacts on human relations has the strongest and most statistically significant effect, such that the greater the negative concern, the more likely that the respondent belongs to Class 2. We find that the other two latent variables too exert smaller and less statistically significant effects on class membership. For example, individuals that view positive impacts to productivity from working from home are more likely to belong to Class 4 over Class 2, and individuals that view positive impacts to health and wellbeing are more likely to belong to Class 3 over Class 2.

However, reducing the measurement indicators to a smaller number of latent factors does lead to some loss of richness in terms of the findings. For example, all indicators loading onto the same factor are now constrained to have the same directional impact on class membership. Whereas in Models 1 and 2, we were able to pick up some differences. For example, accounting for differences in other health and wellbeing indicators, we observed per Model 2 that respondents who believe they are likely to have higher morale are more likely to belong to Class 2, but respondents who believe they are likely to have better work-life balance are *less* likely to belong to Class 2. The present model constrains the effects of each of these indicators to be in the same relative direction (positive or negative, but it cannot be different), where such indicators are collapsed into a single composite latent construct. A reader who has experience with hybrid choice models will already note that the same applies there too.

This distinction reflects a deeper methodological divide between confirmatory and exploratory approaches to modelling attitudes. In psychometrics and related disciplines, confirmatory factor models dominate. Latent constructs are carefully theorized, and measurement indicators are designed to be highly internally consistent, often bordering on redundancy. This ensures reliability and construct validity, and helps guard against responses being overly influenced by the wording of individual items. These are important strengths of the factor-analytic approach. However, in transport and other applied social sciences where hybrid and ICLV models have gained popularity, data collection is often more exploratory. There may be little prior consensus on how indicators should be grouped, or on the theoretical structure of the attitudinal space. In such settings, exploratory approaches like the FMNL or Model 2 offer clear advantages: they allow the analyst to empirically test for distinct effects of individual indicators and uncover nuanced associations between attitudes and behaviours that may not conform to rigid latent structures. This flexibility is especially valuable in behavioural choice contexts, where the goal is not only to validate latent constructs but to understand how different attitudinal dimensions, however

**Table 6:** Class membership model when the factor scores are included as observable explanatory variables (Model 3)

| Variable | Class 1 (reference) | | Class 2 | | Class 3 | | Class 4 | |
|---|---|---|---|---|---|---|---|---|
| | est. | p-value | est. | p-value | est. | p-value | est. | p-value |
| Class-specific constant | 0.000 | - | -0.494 | 0.00 | -0.196 | 0.26 | -0.517 | 0.02 |
| *Attitudinal characteristics* | | | | | | | | |
| Perceived positive impacts on productivity | 0.000 | - | 0.094 | 0.66 | -0.007 | 0.98 | 0.322 | 0.13 |
| Perceived positive impacts on health and wellbeing | 0.000 | - | 0.300 | 0.14 | 0.598 | 0.01 | 0.672 | 0.00 |
| Perceived negative impacts on human relations | 0.000 | - | 1.035 | 0.00 | -0.415 | 0.01 | -0.637 | 0.00 |

**Table 7:** Comparison between mean factor scores denoting different attitudinal constructs across classes, applying posterior profiling to the baseline LCCM

| Attitudinal construct | Mean score | | | | ANOVA F-stat [a] | t-stat | | | | | |
|---|---|---|---|---|---|---|---|---|---|---|---|
| | | | | | | Class 1 v. | | | Class 2 v. | | Class 3 v. |
| | Class 1 | Class 2 | Class 3 | Class 4 | | Class 2 | Class 3 | Class 4 | Class 3 | Class 4 | Class 4 |
| Perceived positive impacts on productivity | -0.201 | 0.082 | -0.019 | 0.216 | 11.91 | 3.83 | -2.67 | -5.63 | 1.47 | -1.80 | 3.42 |
| Perceived positive impacts on health and wellbeing | -0.242 | 0.072 | 0.014 | 0.244 | 14.38 | 4.09 | -3.54 | -6.31 | 0.80 | -2.24 | 3.18 |
| Perceived negative impacts on human relations | 0.014 | 0.287 | 0.141 | -0.400 | 33.95 | 6.18 | 2.02 | 3.70 | 7.94 | 8.75 | -1.96 |

[a] *For an F-distribution with (3, 992) degrees of freedom, if F > 2.60, p-value < 0.05; if F > 3.84, p-value < 0.01; and if F > 6.68, p-value < 0.001*

subtle, shape decision-making. From this perspective, the ability of the FMNL (and Model 2) to retain the specificity of individual indicators is not a methodological limitation, but a substantive strength.

We estimated two additional specifications to test the robustness of the findings. The first was a FMNL model in which the posterior class membership probabilities from the baseline LCCM were regressed on the factor scores. The second was a sequential LCCM where the class-specific choice parameters were fixed to those from the baseline LCCM, and only the class membership model was re-estimated using factor scores. Both models yielded results that were nearly identical to those from Model 3 in terms of sign, magnitude, and significance of parameter estimates. For brevity, we do not report detailed estimation results here, but these supplementary models further reinforce the consistency of the observed associations between latent attitudes and class membership.

We also applied the posterior profiling approach to the factor scores derived from our exploratory factor analysis, comparing mean values across the four latent classes (**Table 7**). A one-way ANOVA test confirmed statistically significant differences in the mean values of all three latent constructs across classes at the 0.001 level. However, the relative magnitude of the F-statistics reveals that not all constructs contribute equally to class differentiation. The F-statistics for the constructs denoting impacts on productivity and health/wellbeing were considerably lower than that for human relations, suggesting that the latter plays a more prominent role in distinguishing attitudinal classes. This pattern is echoed in the pairwise comparisons: between Classes 1, 3, and 4, Class 1 consistently reported lower perceived benefits to productivity and wellbeing. In contrast, Class 4 expressed the fewest concerns about human relations, while Class 3 expressed the most, with Class 1 falling in between. Between Classes 2, 3, and 4, Class 2 reported greater perceived benefits to productivity and wellbeing than Class 3 but fewer than Class 4. However, Class 2 had the highest level of concern about the potential negative impacts on human relations, clearly distinguishing it from both Classes 3 and 4.

Compared to the FMNL approach, where posterior class membership probabilities are explained in terms of factor scores, and the nearly equivalent Model 3, where the factor scores are included as explanatory variables in the class membership model, the posterior profiling approach identifies a broader range of differences across classes. However, because it relies on univariate comparisons, it does not account for the confounding influence of other attitudinal constructs. As a result, it may overstate the significance of some observed differences or fail to isolate the most salient predictors of class membership. In contrast, the FMNL model (and Model 3) provides a more rigorous multivariate assessment that can account for intercorrelations among constructs and reveal which attitudinal dimensions have the strongest independent association with behavioural segmentation. This underscores the value of using both approaches in tandem: posterior profiling offers intuitive and transparent summaries of class-level attitudinal patterns, while the FMNL approach allows for more structured statistical comparison.

### 3.4. Model 4: Hybrid Choice Model

We run a fully specified latent class latent variable hybrid choice model (Model 4; right panel of Figure 2), where the latent variables are loaded on the indicators as before, and included as explanatory variables in the class membership model, and the full model is estimated simultaneously. The estimation results for the model are reported in Appendix B; please note that results for the class-specific choice models were found to be nearly identical to the baseline LCCM.

The findings are consistent with (and almost identical to) those from Model 3, and for the sake of brevity, we do not describe them here again. While the simultaneous estimation of the hybrid model (Model 4) is often promoted as the more theoretically rigorous approach, we find no meaningful difference in parameter estimates or overall model fit compared to the sequential estimation of Model 3. These findings are consistent with previous studies comparing simultaneous and sequential estimation approaches (e.g., Raveau et al., 2010; Bahamonde-Birke & de Dios Ortúzar, 2014), which similarly report negligible differences in results across the two methods. In our case, both models produced near-identical coefficients, statistical significance levels, and behavioural insights, suggesting that the additional complexity of full information maximum likelihood does not translate into practical improvements in explanatory power.

It is often argued that one of the advantages of simultaneous estimation lies in its efficiency. By jointly estimating the measurement and choice components, the model can theoretically achieve tighter standard errors on estimated parameters (Vij and Walker, 2016). In practice, however, we find this benefit to be largely theoretical. Compared to Model 3, the patterns of statistical significance in key parameters remained unchanged. That is, even where the standard errors were marginally reduced in Model 4, this did not affect the outcome of any statistical tests or alter the inferences drawn from the results. Thus, from a hypothesis-testing perspective, the efficiency gains offered by simultaneous estimation appear to have little real impact.

Moreover, the estimation of Model 4 entails a substantial computational burden, reflecting the well-documented complexity of hybrid choice models. These models require the joint estimation of discrete choice and latent variable components, which can lead to demanding likelihood functions, sensitivity to starting values, and non-trivial convergence behaviour - symptoms that are well documented in the literature as endemic to ICLV models (Bolduc and Daziano, 2010; Bhat and Dubey, 2014; Sohn, 2017)[3].

Ultimately, the ICLV framework creates a high methodological bar that analysts are expected to scale in order to claim robustness in modelling attitudes, yet our findings suggest that this bar may be unnecessarily high. For our case study, the full-information hybrid model provided no meaningful improvement over the simpler Model 3, while imposing a significant cost in terms of complexity, transparency, and estimation stability. In practical terms, Model 3 would have led to identical policy conclusions and behavioural interpretations. If the promise of ICLV models lies in rigour, then the challenge for the field is to ensure that this rigour translates into value, not merely difficulty.

### 3.5 Conclusions

Our analysis compared four different frameworks for examining the relationship between attitudes and preferences for working from home (WfH). Both the posterior profiling and FMNL approaches emerged as behaviourally rich and transparent methods, offering complementary strengths: the former allows for intuitive descriptive analysis without structural assumptions, while the latter provides accounts for multivariate association. Model 2, which directly included indicators in a simultaneously estimated class membership model, yielded similar results to both despite theoretical concerns around endogeneity. The use of factor scores in Model 3 brought the structure closer to conventional latent variable models, but at the cost of explanatory richness, since indicators were constrained to act uniformly within each latent construct. Model 4, the fully specified ICLV framework, is often presented as the gold standard for integrating attitudes into choice models. However, in our case, it offered no meaningful improvement in fit, explanatory power, or statistical inference over Model 3, and suffered from similar limitations in terms of loss in explanatory richness.

---

[3] In both case studies, the ICLV specifications were developed following standard practice in the literature. The measurement models were specified using the same sets of attitudinal indicators as in the exploratory factor analysis, with constructs defined based on the factor structures identified in that step. Standard normalisations were applied to ensure identification. In our final implementations, the models converged reliably once appropriately specified.

## 4. Case Study 2: Individual Preferences for COVID-19 Vaccines

Our second case study applies the proposed framework to explore how individual preferences for COVID-19 vaccines vary as a function of attitudinal dispositions such as pandemic-related concern and beliefs about vaccine safety. The data used in this analysis comes from the UK subset of a large, multi-country survey conducted between August and September 2020 and reported in Hess et al. (2022). The survey was designed to examine public attitudes toward COVID-19 vaccination across diverse national contexts, with a focus on understanding the psychological and social factors underlying vaccine acceptance or hesitancy.

The survey was administered online using quota-based sampling to ensure representativeness along key demographic characteristics such as age, gender, and region. In the UK sample, a total of 2,335 adults aged 18 and above participated in the survey. Respondents were presented with a range of questions covering demographic and socio-economic background, political orientation, trust in public institutions, and experiences with COVID-19. In addition, the survey included a series of Likert-scale statements designed to measure attitudinal constructs such as perceived risk of COVID-19 infection, concerns about vaccine safety, trust in vaccine information sources, and beliefs about collective responsibility.

To elicit stated preferences for vaccination, respondents were asked to complete a series of SP experiments comprising hypothetical vaccine scenarios. Each respondent was shown six distinct SP scenarios, such as the example shown in **Figure 4**, where they were offered a choice between free and paid versions of two different vaccines that vary in terms of attributes such as efficacy, risk of side effects, waiting times, and impacts on international travel (for the full list of attributes and levels, please refer to **Table 8**). They were also allowed to choose the option of not being vaccinated. Each scenario thus involved the choice between five possible options, namely free or paid versions of either of the two vaccines, and the option of not being vaccinated.

For the purposes of our analysis, we focus on 2,147 respondents who expressed at least some willingness to consider vaccination, either in the past or in the future, to avoid extreme non-compensatory decision rules. We estimated a series of latent class choice models (LCCMs) with different model specifications, where we varied the explanatory variables, the functional form of the utilities, and the number of classes. Based both on statistical measures of fit and behavioral interpretation, the three-class LCCM was selected as the preferred model specification. To capture potentially greater substitution between the different vaccine options than switching between vaccine and no vaccine, the discrete choice model in each class was of the Nested Logit (NL) type (cf. Train, 2009, chapter 4), grouping together the vaccine options into one nest. Estimation made use of incremental specification building to mitigate local maxima concerns, and the final models were found to be stable across runs. Full estimation details, including model diagnostics and sensitivity analyses, are provided in the original study and its supplementary material (Hess et al., 2022).

The final three-class model specification has a McFadden's adjusted R-squared of 0.289, indicating reasonable goodness-of-fit. For the sake of model parsimony, the class membership model was specified as a constants-only model, and did not include any demographic characteristics as explanatory variables. The class-specific choice models included each of the attributes shown in the SP experiments as the explanatory variables $\mathbf{x}_{\mathbf{ntj}}$, along with a nesting coefficient. Corresponding estimates for the model parameters $\boldsymbol{\beta}_{\mathbf{c}}$ are enumerated in **Table 9**, and a summary of the classes in terms of their shares in the sample population and their aggregated preferences are reported in **Table 10**.

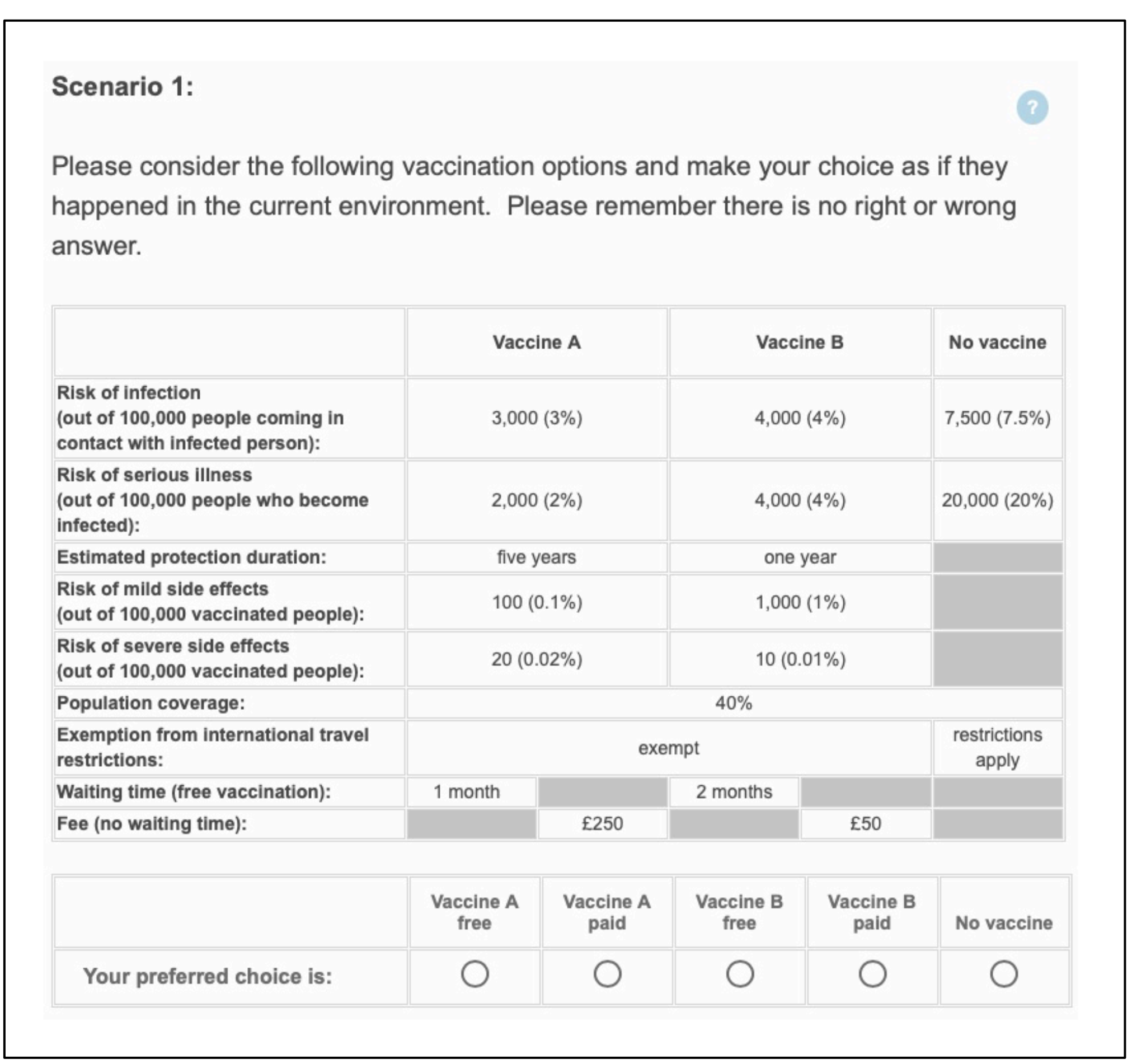

**Scenario 1:**

Please consider the following vaccination options and make your choice as if they happened in the current environment.  Please remember there is no right or wrong answer.

| | Vaccine A | | Vaccine B | | No vaccine |
|---|---|---|---|---|---|
| **Risk of infection (out of 100,000 people coming in contact with infected person):** | 3,000 (3%) | | 4,000 (4%) | | 7,500 (7.5%) |
| **Risk of serious illness (out of 100,000 people who become infected):** | 2,000 (2%) | | 4,000 (4%) | | 20,000 (20%) |
| **Estimated protection duration:** | five years | | one year | | |
| **Risk of mild side effects (out of 100,000 vaccinated people):** | 100 (0.1%) | | 1,000 (1%) | | |
| **Risk of severe side effects (out of 100,000 vaccinated people):** | 20 (0.02%) | | 10 (0.01%) | | |
| **Population coverage:** | 40% | | | | |
| **Exemption from international travel restrictions:** | exempt | | | | restrictions apply |
| **Waiting time (free vaccination):** | 1 month | | 2 months | | |
| **Fee (no waiting time):** | | £250 | | £50 | |

| | Vaccine A free | Vaccine A paid | Vaccine B free | Vaccine B paid | No vaccine |
|---|---|---|---|---|---|
| **Your preferred choice is:** | ○ | ○ | ○ | ○ | ○ |

**Figure 4:** Example screenshot of hypothetical stated preference (SP) scenario to elicit citizen preferences for different COVID-19 vaccines

**Table 8:** Range of attribute values used in our stated preference (SP) experiments to describe different COVID-19 vaccines across different scenarios

| Attribute | Potential values for different COVID-19 vaccines options | | | | | | Value for no vaccine option |
|---|---|---|---|---|---|---|---|
| | Level 1 | Level 2 | Level 3 | Level 4 | Level 5 | Level 6 | |
| Risk of infection out of 100,000 people | 500 (0.5%) | 1,500 (1.5%) | 3,000 (3.0%) | 4,000 (4.0%) | 5,000 (5.0%) | - | 7,500 (7.5%) |
| Risk of illness out of 100,000 people | 2,000 (2%) | 4,000 (4%) | 6,000 (6%) | 10,000 (10%) | 15,000 (15%) | - | 20,000 (20%) |
| Estimated protection duration | 5 years | 2 years | 1 year | 6 months | Unknown | - | - |
| Population coverage | > 80% | 60% | 40% | 20% | < 10% | - | - |
| Risk of mild side effects out of 100,000 people | 100 (0.1%) | 500 (0.5%) | 1,000 (1%) | 5,000 (5%) | 10,000 (10%) | - | - |
| Risk of severe side effects out of 100,000 people | 1 (0.001%) | 5 (0.005%) | 10 (0.010%) | 15 (0.015%) | 20 (0.020%) | - | - |
| Exemption from international travel restrictions | no restrictions | no exemptions | - | - | - | - | Restrictions on international travel |
| Waiting time (for free option) | 2 weeks | 1 months | 2 months | 3 months | 6 months | - | - |
| Cost (GBP) | £10 | £50 | £100 | £200 | £250 | £400 | |

**Table 9:** Class-specific choice models of citizen preferences for COVID-19 vaccines

| Variable | Class 1 | | Class 2 | | Class 3 | |
|---|---|---|---|---|---|---|
| | est. | p-value | est. | p-value | est. | p-value |
| *Alternative specific constants* | | | | | | |
| Alternative shown on left (i.e. Vaccine A) | 0.035 | 0.01 | 0.013 | 0.36 | 0.050 | 0.21 |
| Vaccine is free | 1.452 | 0.00 | 1.066 | 0.00 | -2.935 | 0.00 |
| Vaccine is paid | 1.716 | 0.00 | 0.643 | 0.01 | -3.852 | 0.00 |
| No vaccine (ref.) | 0.000 | - | 0.000 | - | 0.000 | - |
| *Vaccine attributes* | | | | | | |
| Risk of infection out of 100,000 people | -0.153 | 0.00 | -0.125 | 0.00 | -0.142 | 0.00 |
| Risk of illness out of 100,000 people | -0.083 | 0.00 | -0.126 | 0.00 | -0.094 | 0.00 |
| Estimated protection duration (years) | 0.014 | 0.00 | 0.018 | 0.00 | 0.019 | 0.00 |
| Unknown protection duration [1] | -0.390 | 0.00 | -0.291 | 0.00 | 0.000 | - |
| Population coverage | 0.009 | 0.12 | 0.019 | 0.03 | 0.011 | 0.00 |
| Risk of mild side effects out of 100,000 people | -0.052 | 0.00 | -0.042 | 0.00 | -0.050 | 0.00 |
| Risk of severe side effects out of 100,000 people | -16.785 | 0.00 | -21.616 | 0.00 | -33.997 | 0.00 |
| Exemption from international travel restrictions [2] | 0.000 | - | 0.000 | - | 0.174 | 0.48 |
| Waiting time (for free options) | -0.053 | 0.00 | -0.031 | 0.00 | -0.018 | 0.06 |
| Cost (for paid options) | -0.003 | 0.00 | -0.025 | 0.00 | -0.002 | 0.02 |
| Inclusive value (IV) parameter (vaccine nest) [3] | 0.558 | 0.00 | 0.748 | 0.01 | 0.608 | 0.00 |

*[1] Parameter constrained to be negative*
*[2] Parameter constrained to be positive*
*[3] P-value reported for null hypothesis that parameter equals one, alternative hypothesis that parameter is less than one*

**Table 10:** Summary statistics of 3-class model of citizen preferences for COVID-19 vaccines

| Attribute | Class 1 | Class 2 | Class 3 |
|---|---|---|---|
| Share of the sample population | 38.3% | 53.0% | 8.7% |
| Average predicted probability of choosing the following option across different scenarios | | | |
| Free vaccine | 34.6% | 87.5% | 43.9% |
| Paid vaccine | 62.6% | 9.8% | 7.7% |
| No vaccine | 2.8% | 2.7% | 48.4% |

The class-specific estimates reveal clear behavioural segmentation across the sample. Class 3 is strongly resistant to vaccination, and far more likely to opt out across scenarios regardless of vaccine characteristics. By contrast, Classes 1 and 2 display clear preferences for vaccination, but differ in how they respond to cost. Class 1 exhibits a willingness to pay for vaccines, showing only modest sensitivity to price, while Class 2 strongly prefers the free option and is more price-sensitive. Differences in marginal sensitivities to other attributes, such as efficacy or side-effect risks, are less pronounced across the two pro-vaccine classes, suggesting that cost is the primary differentiating factor in their decision-making.

### 4.1 Model 1: Posterior Profiling and the FMNL model

The survey instrument collected responses to a number of Likert-scale statements seeking to measure attitudes towards COVID-19 and vaccine risks. We apply our proposed framework to examine if and how these attitudes and perceptions vary across the three classes. **Table 11** compares the mean responses to the indicators across different classes.

A one-way ANOVA test was conducted to assess whether mean responses to attitudinal indicators varied significantly across the three latent classes identified in the COVID-19 vaccine case study. The results revealed substantial heterogeneity, with several indicators exhibiting highly significant between-class differences. The most discriminating indicators, ranked by the magnitude of their F-statistics, were: "There are significant risks in rapidly developing a vaccine for COVID-19", "I am deeply concerned about COVID-19", and the two opposing statements about government-imposed restrictions: "I believe the measures put in place by the government to restrict transmission need to be strengthened" and "should be relaxed". The statement "I am not sure there will ever be a vaccine" also yielded a high F-statistic. In contrast, some indicators demonstrated very weak or insignificant differences across classes, such as "I believe we will have to live with COVID-19 for a long time", "I am more likely to take risks than others", and concerns about mental wellbeing or economic impacts. These findings indicate that the most salient sources of attitudinal heterogeneity relate to vaccine skepticism and broader perceptions of COVID-19 risk and policy response.

We examine differences between different subsets of classes, using the pairwise t-test statistics. First, we compare Class 3 to Classes 1 and 2. We find that Class 3 is less concerned about COVID-19 in general, but more concerned about its impacts on their personal freedoms and, to a lesser extent, their mental wellbeing. They are more likely to believe that government measures to restrict transmission should be relaxed, and more likely to believe that the risks of vaccination outweigh the benefits. Next, we compare Classes 1 and 2. Class 1 is more likely to be concerned about COVID-19 in general, and its economic effects in particular. Class 1 also sees fewer risks to rapid vaccine development efforts, and is generally more optimistic about efforts to eradicate the disease. However, Class 2 is more likely to believe that healthcare should be free for all, and this likely explains their strong preference for the free vaccine option in the SP experiments.

We run an FMNL model in which the posterior class membership probabilities from the baseline three-class LCCM serve as the dependent variable, and the attitudinal indicators are used as explanatory variables. The estimation results are reported in **Table 12**. As before, we begin by comparing Class 1 with Classes 2 and 3. Consistent with our previous findings, we observe that Class 3 is less concerned about COVID-19 in general, but more concerned about its impacts on their personal freedoms, and less likely to believe that government measures to restrict transmission should be strengthened. They are more likely to believe that the risks of vaccination outweigh the benefits, more likely to agree that rapid vaccine development efforts pose significant risks, and more likely to say they are “not sure there will ever be a vaccine”, confirming a strong degree of vaccine scepticism.

**Table 11:** Comparison between mean responses to different attitudinal statements across classes using the baseline LCCM

| **Attitudinal Measure** *(Level of agreement with statements about self: 1 – strongly disagree, 5 – strongly agree)* | **Mean value** | | | **ANOVA F-stat** [a] | | **t-stat** | | | | | |
|---|---|---|---|---|---|---|---|---|---|---|---|
| | **Class 1** | **Class 2** | **Class 3** | | | **Class 1 v.** | | | | **Class 2 v.** | |
| | | | | | | **Class 2** | | **Class 3** | | **Class 3** | |
| I am deeply concerned about COVID-19 | 4.14 | 3.95 | 3.49 | 28.64 | *** | 4.19 | *** | 7.02 | *** | 5.07 | *** |
| I believe the measures put in place by the government to restrict transmission need to be strengthened | 3.98 | 3.89 | 3.41 | 18.86 | *** | 1.92 | | 5.72 | *** | 4.88 | *** |
| I believe the measures put in place by the government to restrict transmission should be relaxed | 1.88 | 1.94 | 2.40 | 18.74 | *** | 1.44 | | 5.50 | *** | 4.96 | *** |
| I believe that the risks of vaccination outweigh the benefits | 2.34 | 2.33 | 2.84 | 9.95 | *** | 0.07 | | 5.24 | *** | 5.67 | *** |
| There are significant risks in rapidly developing a vaccine for COVID-19 | 3.10 | 3.27 | 3.78 | 34.42 | *** | 3.92 | *** | 8.79 | *** | 6.84 | *** |
| I am concerned about the impact of COVID-19 restrictions on my personal freedoms | 3.03 | 2.99 | 3.48 | 11.89 | *** | 0.76 | | 4.80 | *** | 5.52 | *** |
| I am concerned about the impact of COVID-19 restrictions on my mental wellbeing | 3.17 | 3.10 | 3.31 | 2.67 | | 1.33 | | 1.60 | | 2.50 | * |
| I am concerned about the impact of COVID-19 restrictions on the economy | 4.21 | 4.10 | 4.26 | 4.16 | * | 2.68 | ** | 0.66 | | 2.28 | * |
| I am not sure there will ever be a vaccine | 2.60 | 2.78 | 3.10 | 16.86 | *** | 3.75 | *** | 6.34 | *** | 4.24 | *** |
| I believe we will have to live with COVID-19 for a long time | 4.09 | 4.13 | 4.09 | 0.76 | | 1.25 | | 0.00 | | 0.77 | |
| I am of the opinion that healthcare should be free for all | 4.30 | 4.51 | 4.34 | 12.39 | *** | 5.09 | *** | 0.48 | | 2.53 | * |
| I am more likely to take risks than others | 2.23 | 2.21 | 2.39 | 2.14 | | 0.43 | | 1.87 | | 2.19 | * |

[a] *For an F-distribution with (2, 2144) degrees of freedom, if $F > 3.00$, p-value $< 0.05$; if $F > 4.61$, p-value $< 0.01$; and if $F > 7.00$, p-value $< 0.001$*

** p-value $< 0.05$; ** p-value $< 0.01$; and *** p-value $< 0.001$*

**Table 12:** Fractional logit model when the posterior class membership probabilities from the baseline LCCM are the dependent variables

| Variable | Class 1 (reference) | | Class 2 | | Class 3 | |
|---|---|---|---|---|---|---|
| | est. | p-value | est. | p-value | est. | p-value |
| Class-specific constant | 0.000 | - | 0.202 | 0.68 | -2.371 | 0.00 |
| **Attitudinal Measure** *(Level of agreement with statements about self: 1 – strongly disagree, 7 – strongly agree)* | | | | | | |
| I am deeply concerned about COVID-19 | 0.000 | - | -0.184 | 0.00 | -0.356 | 0.00 |
| I believe the measures put in place by the government to restrict transmission need to be strengthened | 0.000 | - | -0.083 | 0.15 | -0.252 | 0.00 |
| I believe the measures put in place by the government to restrict transmission should be relaxed | 0.000 | - | 0.007 | 0.91 | 0.007 | 0.94 |
| I believe that the risks of vaccination outweigh the benefits | 0.000 | - | -0.023 | 0.44 | 0.116 | 0.01 |
| There are significant risks in rapidly developing a vaccine for COVID-19 | 0.000 | - | 0.163 | 0.00 | 0.610 | 0.00 |
| I am concerned about the impact of COVID-19 restrictions on my personal freedoms | 0.000 | - | -0.005 | 0.91 | 0.147 | 0.05 |
| I am concerned about the impact of COVID-19 restrictions on my mental wellbeing | 0.000 | - | -0.046 | 0.25 | -0.029 | 0.67 |
| I am concerned about the impact of COVID-19 restrictions on the economy | 0.000 | - | -0.128 | 0.01 | -0.113 | 0.23 |
| I am not sure there will ever be a vaccine | 0.000 | - | 0.116 | 0.01 | 0.254 | 0.00 |
| I believe we will have to live with COVID-19 for a long time | 0.000 | - | 0.009 | 0.88 | -0.061 | 0.51 |
| I am of the opinion that healthcare should be free for all | 0.000 | - | 0.270 | 0.00 | 0.142 | 0.10 |
| I am more likely to take risks than others | 0.000 | - | -0.060 | 0.16 | -0.087 | 0.24 |

We observe similar trends as before between Classes 1 and 2. The indicator with the strongest effect on class membership is the statement “I am of the opinion that healthcare should be free for all”, with Class 2 being significantly more likely to agree with the statement. As before, Class 1 is more likely to be concerned about COVID-19 in general, and its economic effects in particular, and Class 1 also sees fewer risks to rapid vaccine development efforts, and is more confident that there will be a vaccine.

In contrast to the first case study, the findings from the FMNL model in this context are strikingly consistent with those obtained from the posterior profiling of class-specific means. This convergence likely reflects the lower degree of multicollinearity among the attitudinal indicators used in the COVID-19 vaccine study, as compared to the highly interrelated indicators in the WfH case. With weaker correlations between indicators, the FMNL model is better able to isolate the marginal contribution of each attitudinal measure to class membership, without the suppression or instability often caused by overlapping explanatory power. As a result, both methods yield a coherent narrative of class-level attitudinal differences: most notably the strong vaccine scepticism and low COVID-19 concern among Class 3, the pro-vaccine but cost-sensitive stance of Class 2, and the high concern about COVID-19 and lower risk perceptions of vaccines exhibited by Class 1.

### 4.2 Model 2: Indicators in the Class Membership Model

Model 2 involves the simultaneous estimation of a latent class choice model in which attitudinal indicators are included directly in the class membership model. This contrasts with the FMNL approach in Section 4.1, which keeps the latent class segmentation fixed. The estimation results for the model are reported in Appendix C. To test robustness, we also estimated a sequential LCCM where the class-specific choice model parameters were held constant, and only the class membership model was re-estimated using the same indicators. As with the previous case study, the two models yielded results nearly identical in sign, magnitude, and significance to the FMNL approach. For the sake of brevity, we do not describe the results in detail again, and we omit estimation results for the sequential model altogether.

As with the first case study, we acknowledge the ongoing debate around potential endogeneity when attitudinal indicators are used directly in the class membership model. However, the empirical consistency of results across all four approaches - posterior profiling, FMNL, Model 2, and the sequential LCCM - suggests that such concerns are not a significant practical issue in this context. This further reinforces the robustness of the underlying attitudinal associations and the stability of the latent class segmentation, regardless of the specific modelling strategy employed.

### 4.3 Model 3: Factor Scores in the Class Membership Model

To explore the latent structure of attitudinal responses, we conducted an exploratory factor analysis on the full set of indicators. Unlike the first case study, where the indicators loaded cleanly onto three distinct and theoretically coherent constructs, the factor analysis here identified only two interpretable factors - the first measures beliefs about the importance of managing risks relating to COVID-19, and the second measures concerns about restrictive measures and their socioeconomic impacts (see **Table 13**)[4]. Several other indicators did not exhibit clear or consistent loadings on any underlying construct and were consequently excluded from the factor-based analysis. As a result, Models 3 and 4 rely on a reduced subset of attitudinal information, potentially limiting their explanatory power.

---

[4] The factor structure is derived using standard exploratory factor analysis, where factors are retained based on statistical criteria such as eigenvalues (e.g. the Kaiser criterion) and the proportion of variance explained, rather than imposed a priori. Under this approach, only two factors met conventional retention thresholds in this dataset, reflecting the relatively weak correlation structure

The class membership model results are listed in **Table 14** (the class-specific models are reported in Appendix C; please note that they were nearly identical to the baseline LCCM). Consistent with previous findings, we observe that individuals belonging to Class 3 are much less likely to support greater containment measures, and much more concerned about the adverse impacts of mobility restrictions, than Classes 1 or 2, though the latter effect is not as statistically significant. Between Classes 1 and 2, Class 1 is more likely to support greater containment measures, as well as more concerned about the adverse impacts of mobility restrictions.

Importantly, since the indicator "I am of the opinion that healthcare should be free for all" did not load on either factor, it is implicitly excluded from Model 3, and consequently offers no insight on differences in class membership. This is a major disadvantage to the approach, as we know from alternative model frameworks that this indicator strongly differentiates Class 1 from Class 2. Similarly, key indicators related to vaccine skepticism, such as "There are significant risks in rapidly developing a vaccine for COVID-19" and "I am not sure there will ever be a vaccine", were also excluded for not loading cleanly onto a single factor. This omission is equally problematic, as our posterior inference framework shows these indicators play a central role in distinguishing Class 3 from the other two. Together, these exclusions highlight how reliance on factor-based dimensionality reduction can suppress important behavioural signals, limiting the explanatory power of the resulting model.

We estimated two additional specifications to test the robustness of the findings. The first was a FMNL model in which the posterior class membership probabilities from the baseline LCCM were regressed on the factor scores. The second was a sequential LCCM where the class-specific choice parameters were fixed to those from the baseline LCCM, and only the class membership model was re-estimated using factor scores. Both models yielded results that were nearly identical to those from Model 3 in terms of sign, magnitude, and significance of parameter estimates. For brevity, we only report estimation results for the FMNL approach (Appendix C), but both supplementary models further reinforce the consistency of the observed associations between latent attitudes and class membership.

We also applied the posterior profiling approach to the factor scores derived from our exploratory factor analysis, comparing mean values across the three latent classes (**Table 15**). A one-way ANOVA test confirmed statistically significant differences across classes for both latent constructs: "Support for risk containment" and "Concern for adverse effects of restrictions", with the F-statistic notably higher for the former, suggesting that support for containment is the stronger attitudinal divider. Next, we conducted pairwise t-tests. Consistent with previous findings, Class 3 is least likely to support greater containment measures, and most likely to be concerned about the adverse impacts of mobility restrictions. As with the preceding factor score-based frameworks, the differences between Classes 1 and 2 are not as clear. Compared to Class 2, Class is both more likely to support

---

among indicators. As such, the resulting aggregation and omission of specific indicators follow directly from the data and the exploratory methodology, rather than from discretionary modelling choices. A confirmatory approach, by contrast, would require a pre-specified measurement structure and is not the focus of the present analysis.

**Table 13:** Factor loadings from exploratory factor analysis of attitudinal indicators

| Attitudinal Indicator | Loadings | |
|---|---|---|
| | Factor 1 | Factor 2 |
| I am deeply concerned about COVID-19 | 0.587 | - |
| I believe the measures put in place by the government to restrict transmission need to be strengthened | 0.825 | - |
| I believe the measures put in place by the government to restrict transmission should be relaxed | -0.756 | - |
| I believe that the risks of vaccination outweigh the benefits | - | - |
| There are significant risks in rapidly developing a vaccine for COVID-19 | - | - |
| I am concerned about the impact of COVID-19 restrictions on my personal freedoms | - | 0.745 |
| I am concerned about the impact of COVID-19 restrictions on my mental wellbeing | - | 0.617 |
| I am concerned about the impact of COVID-19 restrictions on the economy | - | 0.385 |
| I am not sure there will ever be a vaccine | - | - |
| I believe we will have to live with COVID-19 for a long time | - | - |
| I am of the opinion that healthcare should be free for all | - | - |
| I am more likely to take risks than others | -0.325 | - |

**Table 14:** Class membership model when the factor scores are included as observable explanatory variables (Model 3)

| Variable | Class 1 (reference) | | Class 2 | | Class 3 | |
|---|---|---|---|---|---|---|
| | est. | p-value | est. | p-value | est. | p-value |
| Class-specific constant | 0.000 | - | 0.330 | 0.00 | -1.619 | 0.00 |
| *Attitudinal characteristics* | | | | | | |
| Support for risk containment | 0.000 | - | -0.082 | 0.00 | -0.307 | 0.00 |
| Concern for adverse effects of restrictions | 0.000 | - | -0.094 | 0.01 | 0.114 | 0.11 |

**Table 15:** Comparison between mean factor scores denoting different attitudinal constructs across classes, applying posterior profiling to the baseline LCCM

| Attitudinal characteristics | Mean value | | | ANOVA F-stat [a] | t-stat | | |
|---|---|---|---|---|---|---|---|
| | | | | | Class 1 v. | | Class 2 v. |
| | Class 1 | Class 2 | Class 3 | | Class 2 | Class 3 | Class 3 |
| Support for risk containment | 0.21 | 0.01 | -0.96 | 28.66 | 2.44 | 6.10 | 5.09 |
| Concern for adverse effects of restrictions | 0.02 | -0.08 | 0.37 | 9.38 | 1.67 | 3.22 | 4.33 |

[a] *For an F-distribution with (2, 2144) degrees of freedom, if F > 3.00, p-value < 0.05; if F > 4.61, p-value < 0.01; and if F > 7.00, p-value < 0.001*

greater containment measures, and more likely to be concerned about the adverse impacts of mobility restrictions, but the latter difference is statistically weak. As before, several important indicators are absent from the framework.

Unlike in the first case study, where the indicator-based and factor-based approaches were broadly (but not perfectly) aligned, the current results point to an important limitation of relying on latent variables derived through exploratory factor analysis (EFA). Specifically, the EFA process filters out indicators that do not load cleanly onto one of the retained factors. In this case, one such excluded indicator, i.e. "I am of the opinion that healthcare should be free for all", plays a pivotal role in distinguishing between Class 1 and Class 2, two groups that differ subtly but meaningfully in their attitudes toward vaccine cost and accessibility. Similarly, indicators relating to vaccine scepticism, such as "There are significant risks in rapidly developing a vaccine for COVID-19" and "I am not sure there will ever be a vaccine," did not load cleanly onto a single factor and were consequently dropped. These indicators, however, offer strong explanatory value in accounting for differences between Class 3 and the other groups, as shown by our posterior inference framework. By omitting these indicators, the factor-based approach in Models 3 and 4 loses explanatory power and fails to capture the full richness of the attitudinal segmentation. This underscores a key trade-off: while dimensionality reduction can simplify interpretation, it may also blunt the precision of class-level insights when the attitudinal landscape is more nuanced.

It is important to note that the comparison presented here pertains to the *exploratory* use of factor structures, which is the dominant approach in the applied ICLV literature. Under such approaches, factors are derived from the observed correlation structure of the indicators, and the omission or aggregation of specific indicators is an inherent outcome of the factor extraction process rather than a discretionary modelling choice. In settings where correlations between indicators are weak, such as the present case study, this can lead to factor structures that fail to retain distinct attitudinal dimensions captured by individual indicators. By contrast, *confirmatory* factor models, where the measurement structure is specified a priori based on strong theoretical foundations, may yield different results and offer advantages in terms of internal consistency and construct validity. However, such approaches require both well-defined constructs and survey instruments designed to measure them. In more exploratory settings, such as those considered here, retaining indicator-level information can provide additional insight into nuanced attitude–behaviour relationships that may be obscured by aggregation into latent factors.

### 4.4. Model 4: Hybrid Choice Model

We run a fully specified latent class latent variable hybrid choice model (Model 4), in which the latent variables are specified using the same factor structure identified earlier, loaded on the indicators and included as explanatory variables in the class membership model. The full model is estimated simultaneously. The estimation results, shown in Appendix C, are highly consistent with those from Model 3, reaffirming the substantive alignment between sequential and simultaneous estimation approaches. However, as discussed in Section 4.3, the latent variables omit a key indicator that proved critical for distinguishing between Class 1 and Class 2, i.e. level of agreement with the statement "I am of the opinion that healthcare should be free for all". As a result, while the hybrid model provides a coherent and statistically efficient account of class membership, it offers limited additional insight beyond what simpler, sequential estimation methods have already captured. This reinforces our broader conclusion that the value of structural models depends less on their complexity and more on whether their assumptions and data reduction choices are empirically justified.

### 4.5 Conclusions

The second case study reaffirms the utility of posterior inference approaches while highlighting the differences in performance of alternative modelling strategies. When attitudinal indicators are only weakly correlated, as in this case, multivariate approaches like the FMNL model do not suffer from multicollinearity and yield results that are nearly identical to the simpler univariate profiling of class-specific means. At the same time, the FMNL model offers an additional advantage by quantifying the relative importance of different indicators in distinguishing between classes, helping to prioritise which attitudinal dimensions matter most for behavioural segmentation.

However, the same weak correlations imply the absence of a strong underlying factor structure, which complicates dimensionality reduction. As a result, approaches that rely on latent variables, such as posterior profiling of factor scores or hybrid choice models, may inadvertently exclude critical behavioural information embedded in individual indicators. In this instance, indicators central to explaining class distinctions were omitted during factor extraction. For example, the statement "I am of the opinion that healthcare should be free for all" played a crucial role in distinguishing between Classes 1 and 2, while indicators reflecting vaccine scepticism, such as "There are significant risks in rapidly developing a vaccine for COVID-19" and "I am not sure there will ever be a vaccine", were key to differentiating Class 3, yet none of these were retained in the factor structure. Their exclusion weakened the explanatory richness of the factor-based models.

These findings suggest that while more complex models may be appropriate under the right data conditions, simpler methods such as posterior profiling of individual indicators and the FMNL model often provide a more transparent, behaviourally faithful, and robust representation of attitudinal heterogeneity.

## 5. Conclusions

This paper has proposed a pragmatic framework for empirically analysing the relationship between attitudes and behaviour by applying posterior inference methods to latent class choice models (LCCMs). Rather than embedding attitudinal constructs directly within the structural model, a strategy that can lead to interpretive and estimation complexities, we demonstrate how class-specific attitudinal profiles can be recovered through posterior inference, offering a more pragmatic and computationally tractable approach. The framework was applied to two case studies, examining employee preferences for working from home and citizen preferences for COVID-19 vaccines, each drawing on rich attitudinal data and involving diverse attitudinal structures.

Our findings yield several methodological lessons, some of which echo themes already present in the literature, while others offer new perspectives:

First, our proposed posterior inference approaches, both the posterior profiling of class-specific means and the fractional multinomial logit (FMNL) model, yield rich and intuitive insights on the nature of attitude-behaviour relationships, with minimal additional complexity beyond a baseline LCCM. The FMNL model offers a multivariate alternative to univariate profiling, allowing analysts to account for the joint influence of multiple indicators and account for confounding relationships between them. However, when attitudinal indicators are highly collinear, as in the first case study, the FMNL model might only be able to isolate the most important effects. In contrast, univariate posterior profiling can help isolate clean, class-specific differences across all individual indicators, but it lacks the ability to account for potential confounding, and may overstate the behavioural salience of any one indicator. Together, these methods offer complementary perspectives, one accounting for correlation, the other emphasising clarity, each with distinct advantages and limitations depending on the data structure.

Second, dimensionality reduction through exploratory factor analysis can be a double-edged sword. While it offers parsimony and clarity, it may also lead to the exclusion of indicators that are behaviourally meaningful but do not load cleanly onto any latent construct. This trade-off was evident in the second case study, where the omission of indicators significantly weakened the ability of factor-based models to distinguish between key latent classes. In the first case study, the use of factors also obscured important behavioural differences, as the direction of influence was constrained to be the same across indicators loading on the same factor, whereas our posterior inference approaches were able to capture divergent class-level patterns for individual indicators. In contrast to hybrid and factor score models, which require analysts to predefine how indicators group into latent constructs, our posterior inference approach makes no assumptions about the underlying attitudinal structure. This avoids the risk of imposing an ill-fitting or arbitrary factor model and enables more nuanced behavioural insights by retaining attitudinal information at the level of individual indicators.

Third, while the direct inclusion of measurement indicators in the class membership model, particularly in hybrid choice frameworks, has been criticised in the literature for introducing potential endogeneity (Ben-Akiva et al., 2002a, 2002b), our results suggest that such concerns may be overstated in empirical applications. It has been argued that such specifications are misspecified from a causal perspective, as both choices and indicators may be influenced by a common latent construct. While this is a valid concern in some cases, it represents a specific and narrow view of the underlying data-generating process. In many applied settings, it may be entirely reasonable to assume that indicators causally influence choice, particularly when they reflect stable attitudes or prior experiences. Across both case studies, we compared multiple estimation frameworks in which class membership was held fixed in different ways, yet the estimated influence of attitudinal indicators on class membership remained remarkably consistent. All the models examined in this study were based on static, cross-sectional data and can only identify associations, not causal effects, regardless of the modelling framework. Any causal interpretation must therefore be ascribed by the analyst, not inferred from model structure alone. Models with directly included indicators yielded results that were nearly identical to those produced by more behaviourally defensible approaches, providing reassurance that such specifications can still offer valid insights under appropriate conditions.

Finally, the benefits of simultaneous estimation, often promoted as a strength of hybrid modelling frameworks, appear limited in practice. In neither case study did full-information estimation yield results that materially differed from those obtained via sequential estimation or posterior inference approaches. At the same time, the joint estimation of latent variable and choice components introduces greater computational burden, estimation instability, and sensitivity to starting values, issues that are well-documented in the ICLV literature (Bolduc and Daziano, 2010; Bhat and Dubey, 2014; Sohn, 2017). While simultaneous estimation may offer theoretical gains in statistical efficiency (Vij and Walker, 2016), we observed no practical improvements in the precision or significance of estimated parameters. These findings align with prior studies (Raveau et al., 2010; Bahamonde-Birke & de Dios Ortúzar, 2014), which also report negligible empirical differences between simultaneous and sequential estimation.

In summary, our findings suggest that while full-information estimation of ICLV models promises theoretical gains in efficiency and consistency, it sets a high methodological bar that may not always yield commensurate practical benefits in applied transport analysis. Even simpler sequential frameworks that rely on dimensionality reduction methods, such as exploratory factor analysis, suffer from their own shortcomings in terms of interpretability and robustness. The posterior inference methods proposed in this study provide a pragmatic and robust alternative, offering behaviourally meaningful insights without the estimation burden and structural rigidity of hybrid or factor-based models. By disentangling attitudinal and behavioural components, the framework enables transport analysts to identify and interpret preference heterogeneity across behaviourally distinct segments, supporting more targeted policy design, communication strategies, and scenario analysis. Importantly, because these insights are obtained using standard latent class choice models, the approach remains transparent, scalable, and well suited to forecasting and policy evaluation exercises where computational tractability and interpretability are critical.

For example, in the WfH application, our framework allows analysts to distinguish workers who are behaviourally similar in their job choices but differ markedly in their perceptions of productivity, wellbeing, or workplace interactions, informing how flexible work policies or employer messaging might be tailored across segments. Similarly, in the vaccination case study, the framework helps reveal how individuals with comparable choice behaviour differ in underlying risk perceptions and pandemic concern, providing insight into how transport-relevant public health interventions or information campaigns may differentially resonate across the population.

Our framework is not without limitations. First, in both studies examined here, the attitudinal indicators are relatively proximal to the choice context, capturing perceptions and beliefs that directly shape the trade-offs underlying observed behaviour. In such settings, behavioural segmentation combined with posterior inference can recover rich and interpretable attitudinal patterns without requiring fully integrated latent variable structures. However, when the analyst's objective is to represent more distal constructs, such as personal values or broad normative orientations, which influence behaviour only indirectly through intervening attitudes, posterior inference applied to behavioural classes may fail to reveal their role (e.g. Paulssen et al., 2014). In these cases, fully specified ICLV models that explicitly encode hierarchical relationships between latent constructs may remain the more appropriate tool. The framework proposed in this paper should therefore be viewed not as a replacement for ICLV models in all contexts, but as a practical alternative when the primary interest lies in direct attitude–behaviour relationships and transparent behavioural interpretation. However, in those cases, it is then important for analysts to ensure that their findings are not confounded with other sources of heterogeneity. Most applications of hybrid choice models link all random heterogeneity to the attitudinal constructs, which may overstate the role of attitudes, especially if those constructs are broad and less related to the study at hand. When in fact allowing for unrelated random heterogeneity, the role of these constructs in hybrid models may be much smaller than commonly expected (cf. Hess et al., 2018).

Second, and related to the above point, the framework proposed in this paper is explicitly associational in nature. By construction, posterior profiling and the subsequent regression analyses relate estimated class membership patterns to attitudinal indicators without modelling the cognitive or behavioural processes that generate those attitudes. As such, the framework is intended to support behavioural interpretation, diagnostic analysis, and comparative evaluation of modelling strategies in applied settings, rather than to resolve causal concerns. When the research objective is to make causal claims, or to represent complex feedback mechanisms between latent constructs and choice behaviour, fully specified structural approaches, such as ICLV models, remain appropriate. The contribution of the present framework lies instead in offering a transparent, computationally tractable alternative for extracting behavioural insight when such structural modelling is impractical or unnecessary.

Third, we tailored our framework to discrete representations of heterogeneity, as captured by the LCCM. While this structure provides a natural platform for posterior profiling, it is also somewhat bespoke and less general than continuous mixture models such as the mixed logit. An important direction for future research is to adapt our posterior inference approach to settings where preference heterogeneity is modelled as continuous, using posterior distributions of individual-specific coefficients. This would further extend the accessibility and applicability of posterior profiling as a behavioural inference tool across the broader landscape of choice modelling.

**Acknowledgements**

We would like to thank Matt Beck for his suggestion to use ANOVA tests. The WfH dataset was collected with support from iMOVE CRC, funded by the Cooperative Research Centres Program, an Australian Government initiative, as well as the Commonwealth Department of Infrastructure, Transport, Regional Development, Communications and the Arts (DITRDCA), and Transport for New South Wales (TfNSW). Stephane Hess acknowledges the support of the European Research Council through advanced Grant 101020940-SYNERGY.

## Appendix A: Robustness to parameter uncertainty

As discussed in Section 2.2, the posterior class membership probabilities used in our analysis are functions of estimated LCCM parameters and are therefore themselves estimated quantities. The posterior analysis conditions on the maximum likelihood estimates of the baseline LCCM and treats the resulting posterior probabilities as fixed inputs. This simplifies exposition, but does not propagate estimation uncertainty into the posterior profiling statistics.

To assess the robustness of our results to this approximation, we implemented a simulation-based procedure in the spirit of Krinsky and Robb (1986). For each case study, we drew 100 parameter vectors from the estimated asymptotic multivariate normal sampling distribution of the baseline LCCM parameters, using the maximum likelihood point estimates and their estimated covariance matrix. For each draw, we recomputed the posterior class membership probabilities, the corresponding class-specific mean responses to attitudinal indicators, and the associated ANOVA and pairwise t-test statistics reported in the main text.

Table 16 and Table 17 summarise the results for the two case studies, respectively. For each attitudinal statement, the tables report: (i) whether the corresponding ANOVA or pairwise comparison is statistically significant at the 5 per cent level when conditioning on the point estimates of the baseline LCCM, and (ii) the number of times, out of 100 simulated draws, that the same test is statistically significant at the 5 per cent level once first-stage parameter uncertainty is propagated through the posterior probabilities.

Overall, the results indicate that the main conclusions of the paper are highly robust to parameter uncertainty. In both case studies, the ANOVA results were almost always reproduced across the 100 simulations for indicators that were statistically significant in the main text, and were rarely or never significant for indicators that were not. Likewise, most pairwise comparisons that were significant in the main paper remained significant in a large majority, and often all, of the simulated draws. Conversely, pairwise comparisons reported as insignificant in the main paper were generally insignificant across most or all draws.

The few cases in which significance was less stable occurred exclusively for marginal comparisons, that is, comparisons whose F- or t-statistics in the main paper lay very close to the conventional 5 per cent critical value. In these cases, the underlying test statistics already indicate that the differences are borderline, even without reference to p-values. The simulation-based results therefore do not reveal new information, but rather confirm what is already evident from the magnitude of the test statistics themselves: that these effects are weak and should be interpreted with caution. Importantly, these instances do not affect the substantive interpretation of the latent classes or the broader conclusions drawn from the posterior profiling exercise.

Taken together, these results suggest that conditioning on the estimated posterior probabilities in the main analysis does not materially distort the qualitative findings. The posterior profiling framework therefore remains a reliable and informative diagnostic tool.

**Table 16:** Robustness of posterior profiling results to parameter uncertainty - Working-from-home case study

| Indicator | ANOVA F-stat | | | Class 1 v. Class 2 | | | Class 1 v. Class 3 | | | Class 1 v. Class 4 | | | Class 2 v. Class 3 | | | Class 2 v. Class 3 | | | Class 3 v. Class 4 | | |
|---|---|---|---|---|---|---|---|---|---|---|---|---|---|---|---|---|---|---|---|---|---|
| | **S** | **H** | **N** | **S** | **H** | **N** | **S** | **H** | **N** | **S** | **H** | **N** | **S** | **H** | **N** | **S** | **H** | **N** | **S** | **H** | **N** |
| I would be able to focus better on my work | 9.04 | Y | 100 | 4.22 | Y | 100 | 2.87 | Y | 100 | 4.99 | Y | 100 | 2.08 | Y | 77 | 1.21 | N | 1 | 3.26 | Y | 100 |
| I would be able to achieve my job objectives and outputs as expected | 8.70 | Y | 100 | 2.70 | Y | 100 | 4.05 | Y | 100 | 5.46 | Y | 100 | 0.68 | N | 0 | 2.93 | Y | 100 | 2.64 | Y | 100 |
| I would have an increased sense of self-discipline | 6.91 | Y | 100 | 4.07 | Y | 100 | 2.42 | Y | 98 | 4.24 | Y | 100 | 2.33 | Y | 100 | 0.31 | N | 0 | 2.66 | Y | 100 |
| I would be able to multi-task more effectively | 8.60 | Y | 100 | 3.08 | Y | 100 | 1.68 | N | 23 | 5.34 | Y | 100 | 1.94 | N | 50 | 2.43 | Y | 97 | 4.75 | Y | 100 |
| I would have greater life satisfaction | 11.98 | Y | 100 | 3.97 | Y | 100 | 5.23 | Y | 100 | 6.21 | Y | 100 | 0.39 | N | 0 | 2.61 | Y | 100 | 2.44 | Y | 87 |
| I would have higher morale | 12.84 | Y | 100 | 5.53 | Y | 100 | 3.10 | Y | 100 | 5.77 | Y | 100 | 3.27 | Y | 100 | 0.44 | N | 0 | 3.72 | Y | 100 |
| I would have better work-life balance | 13.20 | Y | 100 | 2.57 | Y | 100 | 4.51 | Y | 100 | 7.12 | Y | 100 | 1.35 | N | 3 | 4.84 | Y | 100 | 3.89 | Y | 100 |
| I would experience less stress | 7.48 | Y | 100 | 4.17 | Y | 100 | 2.26 | Y | 85 | 4.52 | Y | 100 | 2.43 | Y | 100 | 0.67 | N | 0 | 3.01 | Y | 100 |
| I would have access to fewer learning opportunities and training sessions | 27.67 | Y | 100 | 5.65 | Y | 100 | 2.55 | Y | 81 | 4.61 | Y | 100 | 8.52 | Y | 100 | 10.17 | Y | 100 | 2.86 | Y | 96 |
| I would be concerned about how my performance would be monitored and observed | 18.57 | Y | 100 | 5.56 | Y | 100 | 1.45 | N | 10 | 2.92 | Y | 100 | 7.21 | Y | 100 | 8.02 | Y | 100 | 1.94 | N | 40 |
| I would be worried that my colleagues are not doing their fair share of the work | 20.11 | Y | 100 | 5.64 | Y | 100 | 2.24 | Y | 67 | 2.75 | Y | 100 | 8.36 | Y | 100 | 8.22 | Y | 100 | 1.14 | N | 1 |
| The relationship with my supervisor would be adversely affected | 29.19 | Y | 100 | 6.52 | Y | 100 | 2.46 | Y | 79 | 3.68 | Y | 100 | 9.36 | Y | 100 | 10.01 | Y | 100 | 1.91 | N | 47 |
| My career prospects may suffer due to loss of ad-hoc interactions with colleagues and supervisors | 14.70 | Y | 100 | 5.15 | Y | 100 | 1.03 | N | 2 | 2.36 | Y | 100 | 6.35 | Y | 100 | 7.11 | Y | 100 | 1.69 | N | 8 |

[a] *For an F-distribution with (3, 992) degrees of freedom, if F > 2.60, p-value < 0.05; if F > 3.84, p-value < 0.01; and if F > 6.68, p-value < 0.001*

[b] S denotes the relevant statistic value reported in the paper conditioned on the point estimates of the baseline LCCM, H denotes if the null hypothesis is rejected at the 5 per cent significance level under this value, and N denotes the number of times, out of 100 simulated draws, that the same test is statistically significant at the 5 per cent level once first-stage parameter uncertainty is propagated through the posterior probabilities

**Table 17:** Robustness of posterior profiling results to parameter uncertainty – COVID-19 vaccine case study

| Indicator | ANOVA F-stat | | | Class 1 v. Class 2 | | | Class 1 v. Class 3 | | | Class 2 v. Class 3 | | |
|---|---|---|---|---|---|---|---|---|---|---|---|---|
| | S | H | N | S | H | N | S | H | N | S | H | N |
| I am deeply concerned about COVID-19 | 28.64 | Y | 100 | 4.19 | Y | 100 | 7.02 | Y | 100 | 5.07 | Y | 100 |
| I believe the measures put in place by the government to restrict transmission need to be strengthened | 18.86 | Y | 100 | 1.92 | N | 40 | 5.72 | Y | 100 | 4.88 | Y | 100 |
| I believe the measures put in place by the government to restrict transmission should be relaxed | 18.74 | Y | 100 | 1.44 | N | 0 | 5.50 | Y | 100 | 4.96 | Y | 100 |
| I believe that the risks of vaccination outweigh the benefits | 9.95 | Y | 100 | 0.07 | N | 0 | 5.24 | Y | 100 | 5.67 | Y | 100 |
| There are significant risks in rapidly developing a vaccine for COVID-19 | 34.42 | Y | 100 | 3.92 | Y | 100 | 8.79 | Y | 100 | 6.84 | Y | 100 |
| I am concerned about the impact of COVID-19 restrictions on my personal freedoms | 11.89 | Y | 100 | 0.76 | N | 0 | 4.80 | Y | 100 | 5.52 | Y | 100 |
| I am concerned about the impact of COVID-19 restrictions on my mental wellbeing | 2.67 | N | 14 | 1.33 | N | 0 | 1.60 | N | 2 | 2.50 | Y | 98 |
| I am concerned about the impact of COVID-19 restrictions on the economy | 4.16 | Y | 100 | 2.68 | Y | 100 | 0.66 | N | 0 | 2.28 | Y | 85 |
| I am not sure there will ever be a vaccine | 16.86 | Y | 100 | 3.75 | Y | 100 | 6.34 | Y | 100 | 4.24 | Y | 100 |
| I believe we will have to live with COVID-19 for a long time | 0.76 | N | 0 | 1.25 | N | 0 | 0.00 | N | 0 | 0.77 | N | 0 |
| I am of the opinion that healthcare should be free for all | 12.39 | Y | 100 | 5.09 | Y | 100 | 0.48 | N | 0 | 2.53 | Y | 100 |
| I am more likely to take risks than others | 2.14 | N | 1 | 0.43 | N | 0 | 1.87 | N | 33 | 2.19 | Y | 85 |

[a] *For an F-distribution with (2, 2144) degrees of freedom, if $F > 3.00$, p-value $< 0.05$; if $F > 4.61$, p-value $< 0.01$; and if $F > 7.00$, p-value $< 0.001$*

[b] S denotes the relevant statistic value reported in the paper conditioned on the point estimates of the baseline LCCM, H denotes if the null hypothesis is rejected at the 5 per cent significance level under this value, and N denotes the number of times, out of 100 simulated draws, that the same test is statistically significant at the 5 per cent level once first-stage parameter uncertainty is propagated through the posterior probabilities

## Appendix B: Estimation Results for Case Study 1

To streamline the presentation in the main text, several supplementary estimation results for Case Study 1 are provided here. These include extended model specifications, alternative formulations, and additional parameter estimates that complement the results discussed in Section 3. While these tables are not essential to following the main narrative, they offer further insight into the robustness of our findings and allow interested readers to explore the underlying detail of our modelling framework.

**Table 18:** Estimation results for class-specific choice models of employee preferences for remote working across different model structures

| Variable | Baseline LCCM | | LCCM with indicators in class membership (Model 2) | | LCCM with factor scores in class membership (Model 3) | | Latent class latent variable model (Model 4) | |
|---|---|---|---|---|---|---|---|---|
| | est. | p-value | est. | p-value | est. | p-value | est. | p-value |
| *Class 1* | | | | | | | | |
| Able to work remotely some days, when possible | 0.317 | 0.19 | 0.290 | 0.12 | 0.302 | 0.13 | 0.340 | 0.10 |
| Able to work remotely some hours, when possible | 0.000 | - | 0.000 | - | 0.000 | - | 0.000 | - |
| Wages ($1,000) | 1.142 | 0.00 | 0.956 | 0.00 | 1.006 | 0.00 | 1.033 | 0.00 |
| *Class 2* | | | | | | | | |
| Able to work remotely some days, when possible | 0.000 | - | 0.000 | - | 0.000 | - | 0.000 | - |
| Able to work remotely some hours, when possible | 0.028 | 0.83 | 0.098 | 0.38 | 0.044 | 0.69 | 0.041 | 0.71 |
| Wages ($1,000) | 0.006 | 0.34 | 0.007 | 0.22 | 0.006 | 0.30 | 0.007 | 0.29 |
| *Class 3* | | | | | | | | |
| Able to work remotely some days, when possible | 2.059 | 0.00 | 2.271 | 0.00 | 2.267 | 0.00 | 2.244 | 0.00 |
| Able to work remotely some hours, when possible | 1.171 | 0.00 | 1.158 | 0.01 | 1.148 | 0.00 | 1.134 | 0.00 |
| Wages ($1,000) | 0.493 | 0.00 | 0.477 | 0.00 | 0.476 | 0.00 | 0.471 | 0.00 |
| *Class 4* | | | | | | | | |
| Able to work remotely some days, when possible | 3.062 | 0.00 | 3.079 | 0.00 | 2.987 | 0.00 | 3.014 | 0.00 |
| Able to work remotely some hours, when possible | 1.475 | 0.00 | 1.430 | 0.00 | 1.494 | 0.00 | 1.501 | 0.00 |
| Wages ($1,000) | 0.119 | 0.00 | 0.116 | 0.00 | 0.114 | 0.00 | 0.114 | 0.00 |

[a] *All class-specific parameters for this case study were constrained to be non-negative*

**Table 19:** Class membership model when the indicators are included as explanatory variables, and the class-specific models are re-estimated simultaneously (Model 2)

| Variable | Measure | Class 1 (reference) | | Class 2 | | Class 3 | | Class 4 | |
|---|---|---|---|---|---|---|---|---|---|
| | *(Level of agreement with statements about self: 1 – strongly disagree, 7 – strongly agree)* | est. | p-value | est. | p-value | est. | p-value | est. | p-value |
| Class-specific constant | NA | 0.000 | - | -4.149 | 0.00 | -1.778 | 0.00 | -2.627 | 0.00 |
| Perceived impacts on productivity | I would be able to focus better on my work | 0.000 | - | 0.142 | 0.22 | -0.071 | 0.60 | -0.016 | 0.90 |
| | I would be able to achieve my job objectives and outputs as expected | 0.000 | - | -0.006 | 0.95 | 0.210 | 0.04 | 0.102 | 0.36 |
| | I would have an increased sense of self-discipline | 0.000 | - | 0.085 | 0.50 | 0.030 | 0.81 | -0.002 | 0.98 |
| | I would be able to multi-task more effectively | 0.000 | - | -0.128 | 0.28 | -0.161 | 0.19 | 0.122 | 0.32 |
| Perceived impacts on health and wellbeing | I would have greater life satisfaction | 0.000 | - | 0.091 | 0.52 | 0.486 | 0.00 | 0.112 | 0.47 |
| | I would have higher morale | 0.000 | - | 0.330 | 0.01 | -0.057 | 0.66 | 0.069 | 0.58 |
| | I would have better work-life balance | 0.000 | - | -0.225 | 0.06 | 0.108 | 0.36 | 0.334 | 0.01 |
| | I would experience less stress | 0.000 | - | 0.014 | 0.89 | -0.117 | 0.28 | -0.047 | 0.65 |
| Perceived impacts on human relations | I would have access to fewer learning opportunities and training sessions | 0.000 | - | 0.091 | 0.28 | -0.149 | 0.14 | -0.235 | 0.01 |
| | I would be concerned about how my performance would be monitored and observed | 0.000 | - | 0.085 | 0.37 | -0.010 | 0.93 | -0.088 | 0.34 |
| | I would be worried that my colleagues are not doing their fair share of the work | 0.000 | - | 0.105 | 0.21 | -0.091 | 0.30 | -0.061 | 0.45 |
| | The relationship with my supervisor would be adversely affected | 0.000 | - | 0.234 | 0.01 | -0.093 | 0.35 | -0.065 | 0.50 |
| | My career prospects may suffer due to loss of ad-hoc interactions with colleagues and supervisors | 0.000 | - | 0.023 | 0.80 | 0.126 | 0.27 | 0.090 | 0.34 |

**Table 20:** Class membership model when the indicators are included as explanatory variables, and the class-specific choice model is constrained to be the same as the baseline LCCM

| **Variable** | **Measure** | **Class 1 (reference)** | | **Class 2** | | **Class 3** | | **Class 4** | |
|---|---|---|---|---|---|---|---|---|---|
| | *(Level of agreement with statements about self: 1 – strongly disagree, 7 – strongly agree)* | **est.** | **p-value** | **est.** | **p-value** | **est.** | **p-value** | **est.** | **p-value** |
| Class-specific constant | NA | 0.000 | - | -3.981 | 0.00 | -1.593 | 0.00 | -2.577 | 0.00 |
| Perceived impacts on productivity | I would be able to focus better on my work | 0.000 | - | 0.133 | 0.23 | -0.054 | 0.64 | -0.009 | 0.94 |
| | I would be able to achieve my job objectives and outputs as expected | 0.000 | - | 0.002 | 0.98 | 0.209 | 0.04 | 0.106 | 0.34 |
| | I would have an increased sense of self-discipline | 0.000 | - | 0.067 | 0.57 | 0.037 | 0.76 | 0.006 | 0.96 |
| | I would be able to multi-task more effectively | 0.000 | - | -0.107 | 0.35 | -0.158 | 0.16 | 0.117 | 0.32 |
| Perceived impacts on health and wellbeing | I would have greater life satisfaction | 0.000 | - | 0.082 | 0.54 | 0.438 | 0.00 | 0.107 | 0.44 |
| | I would have higher morale | 0.000 | - | 0.316 | 0.01 | -0.058 | 0.63 | 0.063 | 0.59 |
| | I would have better work-life balance | 0.000 | - | -0.213 | 0.06 | 0.105 | 0.34 | 0.335 | 0.00 |
| | I would experience less stress | 0.000 | - | 0.009 | 0.93 | -0.107 | 0.32 | -0.046 | 0.65 |
| Perceived impacts on human relations | I would have access to fewer learning opportunities and training sessions | 0.000 | - | 0.104 | 0.20 | -0.134 | 0.18 | -0.235 | 0.01 |
| | I would be concerned about how my performance would be monitored and observed | 0.000 | - | 0.092 | 0.34 | -0.008 | 0.94 | -0.082 | 0.38 |
| | I would be worried that my colleagues are not doing their fair share of the work | 0.000 | - | 0.094 | 0.26 | -0.093 | 0.29 | -0.067 | 0.41 |
| | The relationship with my supervisor would be adversely affected | 0.000 | - | 0.244 | 0.01 | -0.082 | 0.40 | -0.062 | 0.53 |
| | My career prospects may suffer due to loss of ad-hoc interactions with colleagues and supervisors | 0.000 | - | 0.008 | 0.93 | 0.110 | 0.29 | 0.087 | 0.36 |

**Table 21:** Class membership model within a latent class latent variable framework, where the latent variables are included as explanatory variables (Model 4)

| Variable | Class 1 | | Class 2 (reference) | | Class 3 | | Class 4 | |
|---|---|---|---|---|---|---|---|---|
| | **est.** | **p-value** | **est.** | **p-value** | **est.** | **p-value** | **est.** | **p-value** |
| Constant | 0.000 | - | -0.543 | 0.00 | -0.206 | 0.30 | -0.532 | 0.02 |
| *Attitudinal characteristics* | | | | | | | | |
| Perceived positive impacts on productivity | 0.000 | - | 0.065 | 0.76 | -0.213 | 0.32 | 0.222 | 0.26 |
| Perceived positive impacts on health and wellbeing | 0.000 | - | 0.244 | 0.26 | 0.694 | 0.01 | 0.682 | 0.00 |
| Perceived negative impacts on human relations | 0.000 | - | 0.982 | 0.00 | -0.414 | 0.02 | -0.597 | 0.00 |

## Appendix C: Estimation Results for Case Study 2

This appendix compiles supplementary estimation results for Case Study 2 that were relocated from the main text for concision. The tables present additional model estimation results that support the analysis in Section 4. As with Appendix A, these results are intended for readers who wish to examine the modelling outputs in greater detail and to verify the consistency of the findings reported in the main body of the paper.

**Table 22:** Estimation results for class-specific choice models of individual preferences for COVID-19 vaccines across different model structures

| Variable | Baseline LCCM | | LCCM with indicators in class membership (Model 2) | | LCCM with factor scores in class membership (Model 3) | | Latent class latent variable model (Model 4) | |
|---|---|---|---|---|---|---|---|---|
| | est. | p-val | est. | p-val | est. | p-val | est. | p-val |
| Alternative shown on left (i.e. Vaccine A) | | | | | | | | |
| Class 1 | 0.035 | 0.01 | 0.033 | 0.01 | 0.036 | 0.00 | 0.036 | 0.00 |
| Class 2 | 0.013 | 0.36 | 0.013 | 0.31 | 0.013 | 0.36 | 0.013 | 0.36 |
| Class 3 | 0.050 | 0.21 | 0.052 | 0.25 | 0.040 | 0.31 | 0.041 | 0.31 |
| Vaccine is free | | | | | | | | |
| Class 1 | 1.452 | 0.00 | 1.408 | 0.00 | 1.471 | 0.00 | 1.478 | 0.00 |
| Class 2 | 1.066 | 0.00 | 1.097 | 0.00 | 1.098 | 0.00 | 1.098 | 0.00 |
| Class 3 | -2.935 | 0.00 | -3.212 | 0.00 | -2.908 | 0.00 | -2.933 | 0.00 |
| Vaccine is paid | | | | | | | | |
| Class 1 | 1.716 | 0.00 | 1.662 | 0.00 | 1.737 | 0.00 | 1.745 | 0.00 |
| Class 2 | 0.643 | 0.01 | 0.695 | 0.00 | 0.665 | 0.00 | 0.666 | 0.00 |
| Class 3 | -3.852 | 0.00 | -4.227 | 0.00 | -3.799 | 0.00 | -3.826 | 0.00 |
| No vaccine (ref.) | | | | | | | | |
| Class 1 | 0.000 | - | 0.000 | - | 0.000 | - | 0.000 | - |
| Class 2 | 0.000 | - | 0.000 | - | 0.000 | - | 0.000 | - |
| Class 3 | 0.000 | - | 0.000 | - | 0.000 | - | 0.000 | - |
| Risk of infection out of 100,000 people | | | | | | | | |
| Class 1 | -0.153 | 0.00 | -0.150 | 0.00 | -0.154 | 0.00 | -0.154 | 0.00 |
| Class 2 | -0.125 | 0.00 | -0.118 | 0.00 | -0.127 | 0.00 | -0.127 | 0.00 |
| Class 3 | -0.142 | 0.00 | -0.141 | 0.00 | -0.139 | 0.00 | -0.140 | 0.00 |
| Risk of illness out of 100,000 people | | | | | | | | |
| Class 1 | -0.083 | 0.00 | -0.081 | 0.00 | -0.084 | 0.00 | -0.084 | 0.00 |
| Class 2 | -0.126 | 0.00 | -0.119 | 0.00 | -0.127 | 0.00 | -0.128 | 0.00 |
| Class 3 | -0.094 | 0.00 | -0.097 | 0.00 | -0.092 | 0.00 | -0.092 | 0.00 |
| Unknown protection duration [1] | | | | | | | | |
| Class 1 | -0.390 | 0.00 | -0.379 | 0.00 | -0.398 | 0.00 | -0.398 | 0.00 |
| Class 2 | -0.291 | 0.00 | -0.280 | 0.00 | -0.295 | 0.00 | -0.296 | 0.00 |
| Class 3 | 0.000 | - | 0.000 | - | 0.000 | - | 0.000 | - |

[1] *Parameter constrained to be negative*

**Table 23:** Estimation results for class-specific choice models of individual preferences for COVID-19 vaccines across different model structures (contd.)

| Variable | Baseline LCCM | | LCCM with indicators in class membership (Model 2) | | LCCM with factor scores in class membership (Model 3) | | Latent class latent variable model (Model 4) | |
|---|---|---|---|---|---|---|---|---|
| | est. | p-val | est. | p-val | est. | p-val | est. | p-val |
| Estimated protection duration (years) | | | | | | | | |
| Class 1 | 0.014 | 0.00 | 0.013 | 0.00 | 0.014 | 0.00 | 0.014 | 0.00 |
| Class 2 | 0.018 | 0.00 | 0.016 | 0.00 | 0.018 | 0.00 | 0.018 | 0.00 |
| Class 3 | 0.019 | 0.00 | 0.020 | 0.00 | 0.019 | 0.00 | 0.019 | 0.00 |
| Risk of mild side effects out of 100,000 people | | | | | | | | |
| Class 1 | -0.052 | 0.00 | -0.051 | 0.00 | -0.053 | 0.00 | -0.053 | 0.00 |
| Class 2 | -0.042 | 0.00 | -0.040 | 0.00 | -0.042 | 0.00 | -0.042 | 0.00 |
| Class 3 | -0.050 | 0.00 | -0.053 | 0.00 | -0.050 | 0.00 | -0.050 | 0.00 |
| Risk of severe side effects out of 100,000 people | | | | | | | | |
| Class 1 | -16.785 | 0.00 | -16.197 | 0.00 | -17.094 | 0.00 | -17.043 | 0.00 |
| Class 2 | -21.616 | 0.00 | -20.715 | 0.00 | -21.730 | 0.00 | -21.788 | 0.00 |
| Class 3 | -33.997 | 0.00 | -37.152 | 0.00 | -32.963 | 0.00 | -33.539 | 0.00 |
| Waiting time (for free options) | | | | | | | | |
| Class 1 | -0.053 | 0.00 | -0.052 | 0.00 | -0.054 | 0.00 | -0.054 | 0.00 |
| Class 2 | -0.031 | 0.00 | -0.030 | 0.00 | -0.031 | 0.00 | -0.031 | 0.00 |
| Class 3 | -0.018 | 0.06 | -0.015 | 0.13 | -0.018 | 0.04 | -0.019 | 0.04 |
| Cost (for paid options) | | | | | | | | |
| Class 1 | -0.003 | 0.00 | -0.003 | 0.00 | -0.003 | 0.00 | -0.003 | 0.00 |
| Class 2 | -0.025 | 0.00 | -0.024 | 0.00 | -0.025 | 0.00 | -0.025 | 0.00 |
| Class 3 | -0.002 | 0.02 | -0.002 | 0.06 | -0.002 | 0.02 | -0.002 | 0.02 |
| Population coverage | | | | | | | | |
| Class 1 | 0.009 | 0.12 | 0.010 | 0.10 | 0.009 | 0.14 | 0.009 | 0.14 |
| Class 2 | 0.019 | 0.03 | 0.016 | 0.03 | 0.017 | 0.05 | 0.017 | 0.05 |
| Class 3 | 0.011 | 0.00 | 0.011 | 0.00 | 0.011 | 0.00 | 0.011 | 0.00 |
| Exemption from international travel restrictions [2] | | | | | | | | |
| Class 1 | 0.000 | - | 0.000 | - | 0.000 | - | 0.000 | - |
| Class 2 | 0.000 | - | 0.000 | - | 0.000 | - | 0.000 | - |
| Class 3 | 0.174 | 0.48 | 0.254 | 0.37 | 0.204 | 0.39 | 0.211 | 0.39 |
| Inclusive value (IV) parameter (vaccine nest) [3] | | | | | | | | |
| Class 1 | 0.558 | 0.00 | 0.547 | 0.00 | 0.563 | 0.00 | 0.562 | 0.00 |
| Class 2 | 0.748 | 0.00 | 0.703 | 0.00 | 0.755 | 0.00 | 0.757 | 0.00 |
| Class 3 | 0.608 | 0.00 | 0.633 | 0.00 | 0.610 | 0.00 | 0.612 | 0.00 |

[2] *Parameter constrained to be positive*

[3] *P-value reported for null hypothesis that parameter equals one, alternative hypothesis that parameter is less than one*

**Table 24:** Class membership model when the indicators are included as explanatory variables (Model 2)

| Variable | Class 1 (reference) | | Class 2 | | Class 3 | |
|---|---|---|---|---|---|---|
| | est. | p-value | est. | p-value | est. | p-value |
| Class-specific constant | 0.000 | - | 0.180 | 0.75 | -2.717 | 0.02 |
| **Attitudinal Measure** *(Level of agreement with statements about self: 1 – strongly disagree, 7 – strongly agree)* | | | | | | |
| I am deeply concerned about COVID-19 | 0.000 | - | -0.213 | 0.00 | -0.449 | 0.00 |
| I believe the measures put in place by the government to restrict transmission need to be strengthened | 0.000 | - | -0.093 | 0.17 | -0.322 | 0.00 |
| I believe the measures put in place by the government to restrict transmission should be relaxed | 0.000 | - | 0.009 | 0.90 | -0.031 | 0.76 |
| I believe that the risks of vaccination outweigh the benefits | 0.000 | - | -0.026 | 0.48 | 0.141 | 0.01 |
| There are significant risks in rapidly developing a vaccine for COVID-19 | 0.000 | - | 0.179 | 0.00 | 0.858 | 0.00 |
| I am concerned about the impact of COVID-19 restrictions on my personal freedoms | 0.000 | - | -0.009 | 0.87 | 0.178 | 0.08 |
| I am concerned about the impact of COVID-19 restrictions on my mental wellbeing | 0.000 | - | -0.049 | 0.30 | -0.047 | 0.61 |
| I am concerned about the impact of COVID-19 restrictions on the economy | 0.000 | - | -0.150 | 0.01 | -0.130 | 0.29 |
| I am not sure there will ever be a vaccine | 0.000 | - | 0.138 | 0.01 | 0.337 | 0.00 |
| I believe we will have to live with COVID-19 for a long time | 0.000 | - | 0.015 | 0.83 | -0.094 | 0.46 |
| I am of the opinion that healthcare should be free for all | 0.000 | - | 0.317 | 0.00 | 0.127 | 0.30 |
| I am more likely to take risks than others | 0.000 | - | -0.081 | 0.11 | -0.140 | 0.16 |

**Table 25:** FMNL model where the class membership probabilities are regressed on the factor scores

| Variable | Class 1 (reference) | | Class 2 | | Class 3 | |
|---|---|---|---|---|---|---|
| | **est.** | **p-value** | **est.** | **p-value** | **est.** | **p-value** |
| Class-specific constant | 0.000 | - | 0.330 | 0.00 | -1.580 | 0.00 |
| *Attitudinal characteristics* | | | | | | |
| Support for risk containment | 0.000 | - | -0.073 | 0.00 | -0.260 | 0.00 |
| Concern for adverse effects of restrictions | 0.000 | - | -0.081 | 0.02 | 0.091 | 0.12 |

**Table 26:** Class membership model within a latent class latent variable framework, where the latent variables are included as explanatory variables (Model 4)

| Variable | Class 1 (reference) | | Class 2 | | Class 3 | |
|---|---|---|---|---|---|---|
| | **est.** | **p-value** | **est.** | **p-value** | **est.** | **p-value** |
| Class-specific constant | 0.000 | - | 0.331 | 0.00 | -1.644 | 0.00 |
| *Attitudinal characteristics* | | | | | | |
| Support for risk containment | 0.000 | - | -0.184 | 0.00 | -0.680 | 0.00 |
| Concern for adverse effects of restrictions | 0.000 | - | -0.143 | 0.03 | 0.215 | 0.05 |